\documentclass[fleqn,11pt]{wlscirep}
\usepackage[utf8]{inputenc}
\usepackage[T1]{fontenc}
\usepackage[numbers,compress]{natbib}
\usepackage{setspace,bm,lipsum,mwe,graphicx,lipsum,boldline,multicol,booktabs,multirow,setspace,amsmath}
\usepackage[font=small,labelfont=bf,justification=justified]{caption}
\title{Quasi-static responses of marine mussel plaques detached from deformable wet substrates under directional tensions}
\author[1]{Yong Pang}
\author[1,$\dag$]{Tao Liu}
\affil[1]{School of Engineering and Materials Science, Queen Mary University of London, London E1 4NS, UK}
\affil[$\dag$]{Corresponding author(s): Tao Liu, email: \textcolor{blue}{tao.liu@qmul.ac.uk}}

\begin{abstract}
Quantifying the response of marine mussel plaque attachment on wet surfaces remains a significant challenge to a mechanistic understanding of plaque adhesion. Here, we developed a customised microscopy system combined with two-dimensional (2D) in-situ digital image correlation (DIC) to quantify the in-plane deformation of a deformable substrate that interacts with a mussel plaque while under directional tension. By analysing the strain field in the substrate, we gained insight into how in-plane traction forces are transmitted from the mussel plaque to the underlying substrate. Finite element (FE) models were developed to assist the interpretation of the experimental measurement. Our study revealed a synergistic effect of pulling angle and substrate stiffness on plaque detachment, with mussel plaques anchoring to a “stiff” substrate at a smaller pulling angle having mechanical advantages with higher load-bearing capacity and less plaque deformation. We identified two distinct failure modes, i.e., shear traction-governed failure (STGF) mode and normal traction-governed failure (NTGF). It was found that increasing the substrate stiffness or reducing the pulling angle resulted in a failure mode change from NTGF to STGF. Our findings offer new insights into the mechanistic understanding of plaque/substrate interaction, which provides a general plaque-inspired strategy for wet adhesion.
\end{abstract}

\begin{document}
%\doublespacing
\flushbottom
\maketitle
\thispagestyle{empty}
\noindent keywords: \emph{Mytilus edulis}, Mussel plaque, Surface traction, Digital image correlation, Optical microscopy

\section*{Introduction}
\begin{figure*}[!tp]
\centering
\includegraphics[width=0.95 \linewidth]{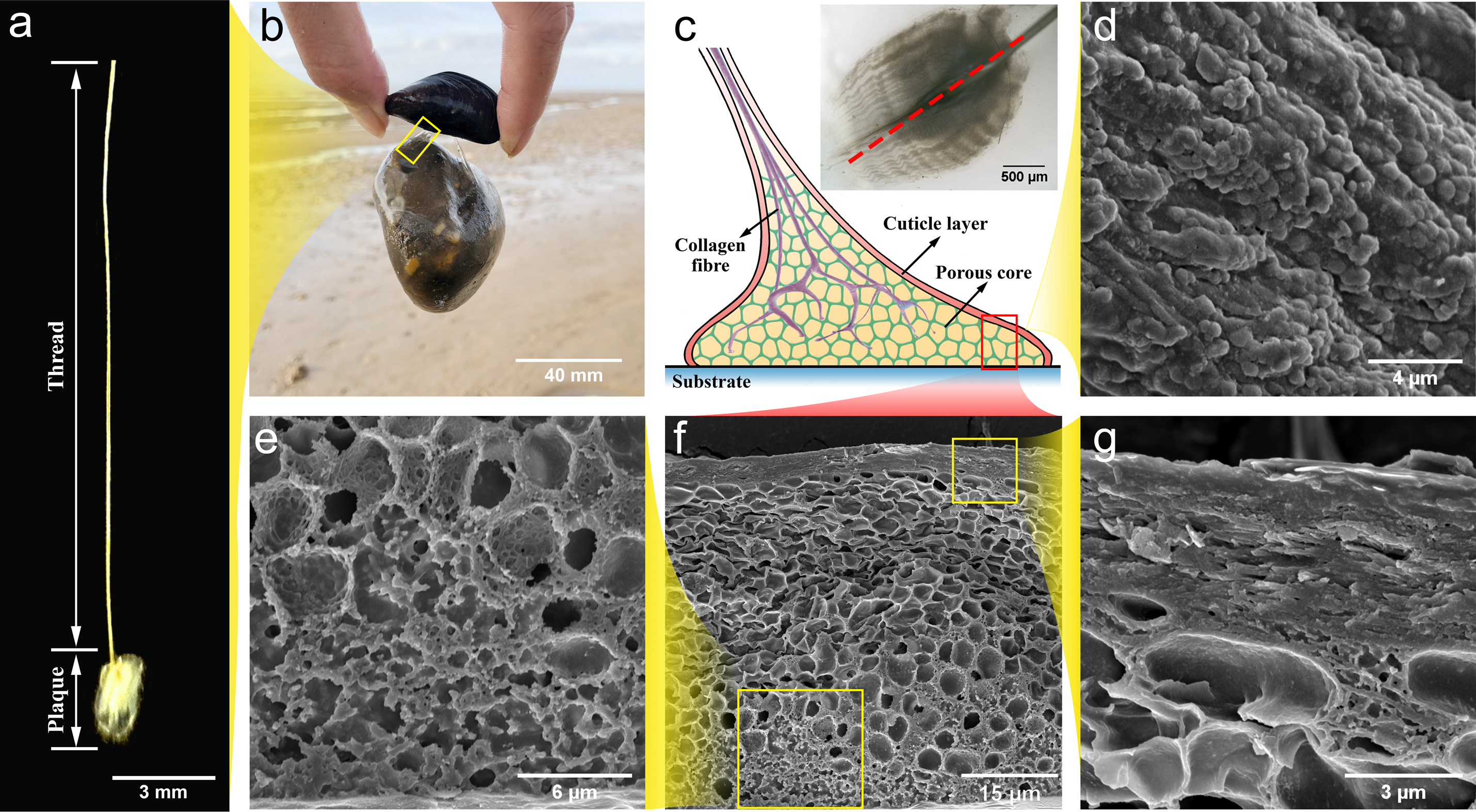}
\caption{\textbf{Characterisation of unique adhesive structure in mussel plaque.} \textbf{a} Single plaque-thread system of marine mussel. \textbf{b} Mussel plaque anchorage system carrying a 248g rock, approximately 6 to 10 times mussel weight. \textbf{c} Schematic of single mussel plaque showing the adhesive structures. The inset shows the top view of a true mussel plaque, and the dashed line shows the section direction for the plaque schematic. \textbf{d} Micrograph of distal thread surface covered by protective granular composite structure. \textbf{e} Micrograph of highly dense pores in the vicinity of plaque/substrate interface. \textbf{f} Overview of the hierarchical adhesive microstructure of mussel plaque. \textbf{g} The cross-section shows the thickness of the cuticle layer, around $4\mu m$.}
\label{Fig.1}
\end{figure*}

Underwater adhesion is a common phenomenon in aquatic organisms, allowing them to survive in harsh natural environments \cite{Fan2021,Stewart2011,Chen2020}. The physical mechanism of underwater adhesion can be achieved at the macroscale or micro/nanoscale via suction \cite{Heydari2020, Green2013, Tramacere2013}, capillarity \cite{Meng2019,Ditsche2019,Lin2009,David2015,David2014}, interlocking \cite{Yang2020,Park2017,Amador2017,Lee2019,Chuang2017,Ditsche2010}, and protein-based chemical bonds. Recently, the adhesive mechanism based on chemical bonds, especially marine mussels, has received significant attention \cite{Kavanagh2001,Rittschof2008,Holm2006,Lin2007}. The plaques of marine mussels are an incredible design of nature, which anchor themselves to wet surfaces in hostile intertidal zones such as rocks, salt-encrusted and corroded surfaces via collagen-rich threads ending in adhesive plaques \cite{George2018,Priemel2021}, as shown in Fig. \ref{Fig.1}a. Radially distributed thread-plaque systems provide strong anchorage up to ten times the self-weight of a single mussel (Fig. \ref{Fig.1}b), enabling them to survive the hydrodynamic forces exerted by tidal currents or the force of predators hoping to dislodge them for a meal.

Recent research progress has suggested that, in addition to the interaction of protein-based chemistry at the adhesion sites \cite{Lin2007,Yu2013,Lee2006}, the unique adhesive structure of a mussel plaque plays a significant role in wet adhesion \cite{Waite2017, Filippidi2015,Qin2013}. The adhesive structure of a mussel plaque includes an outer, dense protective cuticle layer \cite{Valois2019,Jehle2020}, collagen fibre bundles \cite{Harrington2010} and a low-density, porous plaque core \cite{Filippidi2015,Bernstein2020}, as shown in Fig. \ref{Fig.1}c. The plaque core exhibits a porous structure consisting of a disordered, foamy network of pores at the length scales ranging from the nanoscale (90 - 120 nm) to the microscale (1.5 - 2.5 µm) (Fig. \ref{Fig.1}f), which is reminiscent of cellular solids \cite{Chisca2018,Yang2022,Shi2021} or lattice structures \cite{Pham2019,Liu2021,Shaikeea2022,Bhuwal2023,Yang20222} with enhanced mechanical performance and damage-tolerant properties. The cuticle layer is protected by a granular composite structure at the length scale of 0.5 µm (Fig. \ref{Fig.1}d), approximately 4 µm in thickness (Fig. \ref{Fig.1}g). The cuticle layer and the plaque core work cooperatively to achieve superior load-bearing capacity under dynamic \cite{Qin2013, Aldred2007} and static loads \cite{Xu2019, Aldred2007}. Although the microstructures of mussel plaque have been extensively studied \cite{Filippidi2015,Bernstein2020,Rao2022}, available studies have never reported a gradient change in pore size in the vicinity of the plaque/substrate interface (Figs. \ref{Fig.1}e and f). The presence of this pore arrangement may affect the load distribution at the underlying substrates. Therefore, studying deformation patterns at deformable substrates offers fundamental insights into the failure mechanisms at the adhesive interface.

Prior research has identified several factors that affect plaque detachment, including pulling angles \cite{Desmond2015,Qin2013}, substrate surface conditions \cite{Kwon2021,Amini2017,Kang2016}, strain rates \cite{Carrington2004}, and loading cycles \cite{Wilhelm2017,Bertoldi2007,Carrington2004}. Although mechanical responses of a plaque-thread system have been successfully characterised, these experiments were mainly conducted in moist or dry conditions \cite{Desmond2015,Qin2013,Wilhelm2017}, which may not fully represent the failure mechanism of underwater plaque detachment. Additionally, recent studies on the mechanosensing mechanism of mussel feet have established that mussels prefer hard or hydrophilic surfaces for anchoring \cite{Amini2017,Choi2021}. However, the in-depth interpretation of the mechanical advantage of anchoring to these substrates remains unclear. Therefore, investigating the underwater plaque/substrate interaction, such as surface traction, strain heterogeneity at substrates, and plaque deformation, can deepen the mechanistic understanding of the adhesive behaviour of mussel plaque.

The present study investigates the effects of pulling angles and substrate stiffness on plaque detachment and advances our understanding of plaque/substrate interaction in underwater environments. To capture the deformation pattern at substrates, a method was proposed to fabricate deformable polydimethylsiloxane (PDMS) substrates, followed by developing a customised microscopy system to characterise the mechanical behaviours and failure mechanism of plaque detachment. Numerical models were subsequently developed to assist the interpretation of substrate deformation, which was verified by mechanical tests and 2D in-situ digital image correlation (DIC) measurements. By shedding light on the plaque and substrate interaction, this study significantly contributes to the knowledge of marine mussel adhesive mechanisms and provides valuable insights for developing plaque-inspired adhesive structures, with potential applications in artificial underwater anchorage systems.

\section*{Results}
\subsection*{Experimental setup and computational modelling}
\begin{figure*}[!tp]
\centering
\includegraphics[width=0.95 \linewidth]{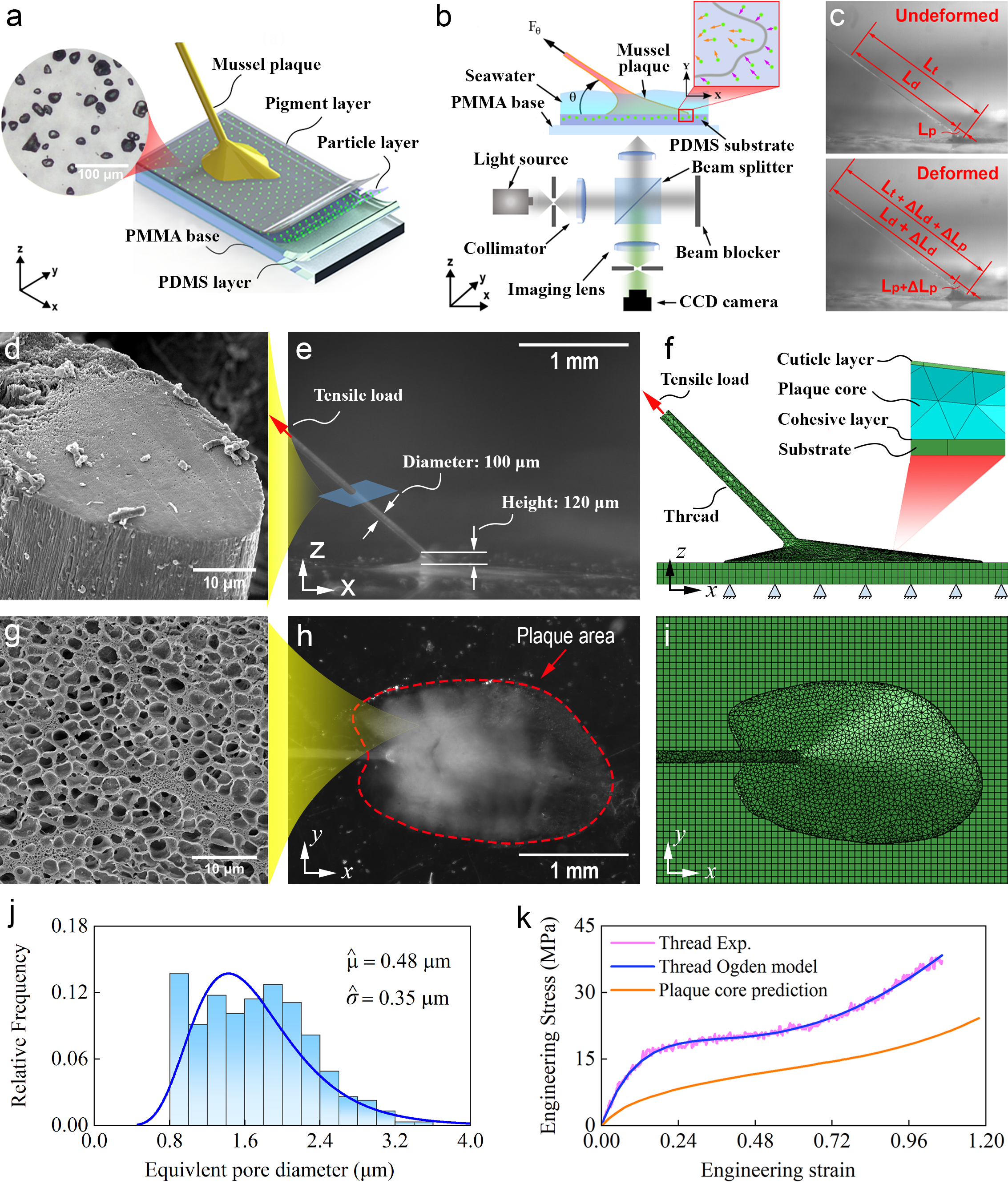}
\caption{\textbf{Experimental and numerical approaches for the characterisation of plaque/substrate interaction.} \textbf{a} Fabrication of four-layer PDMS substrates for mussel plaque deposition. \textbf{b} The schematics of the microscopy system employed in directional tensile tests. \textbf{c} A mussel thread-plaque system under undeformed and deformed configurations. \textbf{d} The cross-section of mussel thread. \textbf{e} and \textbf{f} The side view of a real plaque-thread system and FE model, respectively. \textbf{g} Porous microstructures in mussel plaque core. \textbf{h} The projection of a real mussel plaque on a substrate. \textbf{i} Top view of the FE model. \textbf{j} Pore size distribution of mussel plaque core. \textbf{k} Determination of thread and plaque core properties for FE simulation.}
\label{Fig.2}
\end{figure*}

To track the substrate deformation, 4-layer substrates (Fig. \ref{Fig.2}a) were built layer upon layer using spin coating method. Polymethyl methacrylate (PMMA) plates with a dimension of $60\times60\times2$ mm were employed as the base of PDMS coating. PDMS layer was coated in a thickness of approximately $200 \mu m$, which ensures the mechanical property is thickness independent \cite{Liu2009} and the substrate deformation is measurable. The stiffness of the PDMS layer was tuned from $0.57 \pm  0.01$ MPa (“soft” substrate)  to $1.68 \pm  0.05$ MPa (“stiff” substrate). Micro-particles were mixed with PDMS to create a monolayer of randomly distributed particles for deformation tracking. Black liquid silicone pigment was added into the pigment layer to reduce the reflective lights from the plaque and enhance the imaging contrast of micro-particles. Both the particle and pigment layers were spin-coated with an ultra-thin thickness of approximately 15 $\mu m$, approaching the manufacturing limits of the spin coater (Laurell WS-650-23B). A customised microscopy system was designed to characterise the surface traction, the failure mechanism of plaque detachment and deformation patterns at substrates (Fig. \ref{Fig.2}b and Supplementary Fig. \textcolor{blue}{S4}). During the experiment, mussel plaques were fully immersed in seawater to measure their underwater mechanical responses. Directional tensile loads were applied to the mussel threads at various pulling angles with respect to the substrate, ranging from 15$^\circ$ to 90$^\circ$  at an increment of 15$^\circ$.

Throughout the paper, the global coordinates $x$-$y$-$z$ are defined as follows: the $x$-axis is aligned with the projection of the thread on the substrate; the $z$-axis is perpendicular to the substrate and the $y$-axis is determined by the right-hand rule. Under a directional tensile load $\mathbf{F_\theta}(\mathbf{F_\theta}=F_\theta \mathbf{n})$, the average surface traction ($\mathbf{T}$), and its shear ($T_t$) and normal ($T_n$) components can be defined as
\begin{equation}
\mathbf{T}=\frac{F_\theta}{A}\mathbf{n}=T_t\boldsymbol{e_x}+T_n\boldsymbol{e_z}
\label{Eq1}
\end{equation}
\begin{equation}
T_t=\frac{1}{A}\int_{A}^{}t_tdA=\frac{F_\theta Cos\theta}{A}
\label{Eq2}
\end{equation}
\begin{equation}
T_n=\frac{1}{A}\int_{A}^{}t_ndA=\frac{F_\theta Sin\theta}{A}
\label{Eq3}
\end{equation}
where $\mathbf{n}$ and $F_{\theta}$ denote the direction vector and magnitude of  the tensile force applied to the free end of the mussel thread (Fig. \ref{Fig.2}b), respectively; $\boldsymbol{e_x}$ and $\boldsymbol{e_z}$ the unit vectors in $x$ and $z$-directions; $t_t$ and $t_n$ are the normal ($z$-direction) and tangential ($x$-direction) components of the local surface traction; $\theta$ the pulling angle defined as an angle from the projection of the thread on the substrate to the thread in the clockwise direction (Fig. \ref{Fig.2}b); $A$, the area of the projection of a mussel plaque on the underlying substrate, defined as the area enclosed by the red dashed curve in Fig. \ref{Fig.2}e. The total tensile strains of a thread–plaque system $\boldsymbol{\varepsilon_t} (\boldsymbol{\varepsilon_t} = \varepsilon_t \mathbf{n})$, the strain of the plaque in the system $\boldsymbol{\varepsilon_p} (\boldsymbol{\varepsilon_p} = \varepsilon_p \mathbf{n})$, and the strain of the thread in the system $\boldsymbol{\varepsilon_d} (\boldsymbol{\varepsilon_d} = \varepsilon_d \mathbf{n})$ can be defined as

\begin{equation}
\varepsilon_t=\frac{\Delta L_d+\Delta L_p}{L_t}=\frac{\Delta L_d+\Delta L_p}{L_d+L_p}
\label{Eq4}
\end{equation}
\begin{equation}
\varepsilon_p=\frac{\Delta L_p}{L_p}
\label{Eq5}
\end{equation}
\begin{equation}
\varepsilon_d=\frac{\Delta L_d}{L_d}
\label{Eq6}
\end{equation}
Where $L_t$, $L_d$, and $L_p$ represent the original length of thread–plaque system, the thread and the plaque under undeformed configuration, respectively (Fig. \ref{Fig.2}c), with $L_t=L_d+L_p$; $\Delta L_d$ and $\Delta L_p$ denote the elongations of the thread and the plaque, respectively.

Full-scale, three-dimensional (3D) FE simulations were conducted using the commercially available FE solver \emph{ABAQUS/Explicit}. Each FE model contained five components, a thread, a cuticle layer, a plaque core, a cohesive interface, and a substrate, as shown in Figs. \ref{Fig.2}f and i. Directional tensile loads were applied to the end of the mussel thread using displacement control, as shown in Figs. \ref{Fig.2}e and f. The dimensions of the FE models were chosen in accordance with those of the real mussel plaques (e.g., Figs. \ref{Fig.2}e and h): the thread is approximately 100 µm in diameter; the mussel plaque, a core-shell structure, covered by a 4 µm thick cuticle layer (as measured in Fig. \ref{Fig.1}g), is 120 mm in height. The thread and plaque core were modelled with 4-node 3D tetrahedral elements, i.e., the C3D4 elements in \emph{ABAQUS} notation, and the cuticle layer was modelled with 6-node 3D wedge elements (C3D6 elements). The SEM images suggest that both the thread and the cuticle layer possess a similarly dense, solid microstructure (Fig. \ref{Fig.2}d and Fig. \ref{Fig.1}g). They were assumed to have similar mechanical behaviour and, therefore, modelled by an identical material model. A uniaxial tensile test was conducted to determine the mechanical behaviour of a single thread, in which the strain of the single thread ($\varepsilon_d$) was defined by Eq. \ref{Eq6}. The experimental result of a single thread tension was fitted into the three-order Ogden's model \cite{Ogden1972}, as shown in Fig. \ref{Fig.2}k.

Plaque cores exhibit porous structures with a pore volume fraction of $37.4\% - 50.2\%$ \cite{Filippidi2015,Luke2020,Bernstein2020}. The mechanical property of the plaque cores was estimated based on the FE simulations on the two-dimensional (2D) representative volume element (RVE) under the uniaxial tensile test with periodic boundary conditions \cite{Borys2016}. The RVE was created by mimicking the pore distribution within plaque cores using the approach described by \cite{Cao2019}. The SEM images such as Fig. \ref{Fig.2}g were used to measure the pore distributions, which suggested that the pores followed a log-normal distribution with a mean of the log of the radius $\hat{\mu}=0.48$ µm and a standard deviation of the log of the radius $\hat{\delta}= 0.35$ µm (Fig. \ref{Fig.2}j). The parent material of the RVE was assumed to be identical to that of the cuticle layer. Figure \ref{Fig.2}k shows the predicted mechanical behaviour which is fitted by Ogden’s model (Eq. \ref{Eq7} in the Methods section) using the material data shown in Table \ref{Tab.1}. Numerical tests suggested that the predictions obtained by 3D RVEs did not show significant differences from those obtained by 2D RVEs. The mussel foot proteins (mfps) between the plaque and the substrate, an ultra-thin adhesive layer of less than 15 nm \cite{Valois2020}, were modelled with 6-node three-dimensional cohesive elements (COH3D6 elements). The adhesive behaviour of mfps was modelled by the traction versus separation relation \cite{Camanho2002}. 

The PDMS substrate, with dimensions of 3.2×2.7×0.2 mm, was modelled with 8-node 3D brick elements with a reduced integration scheme (C3D8R elements). Both the numerical experiments and experimental measurements suggested that the tensile loads were insufficient to trigger nonlinear substrate responses. Therefore, the substrates were modelled as linear elastic materials using Young’s modulus of $E=1.68$ MPa for "stiff" PDMS and $E=0.56$ MPa for "soft" PDMS (Supplementary Figs. \textcolor{blue}{S2}b and e) and Poisson’s ratio of 0.49 \cite{Pritchard2013}.

\subsection*{Responses of a thread-plaque system under directional tensions}
\begin{figure*}[!b]
\centering
\includegraphics[width=1 \linewidth]{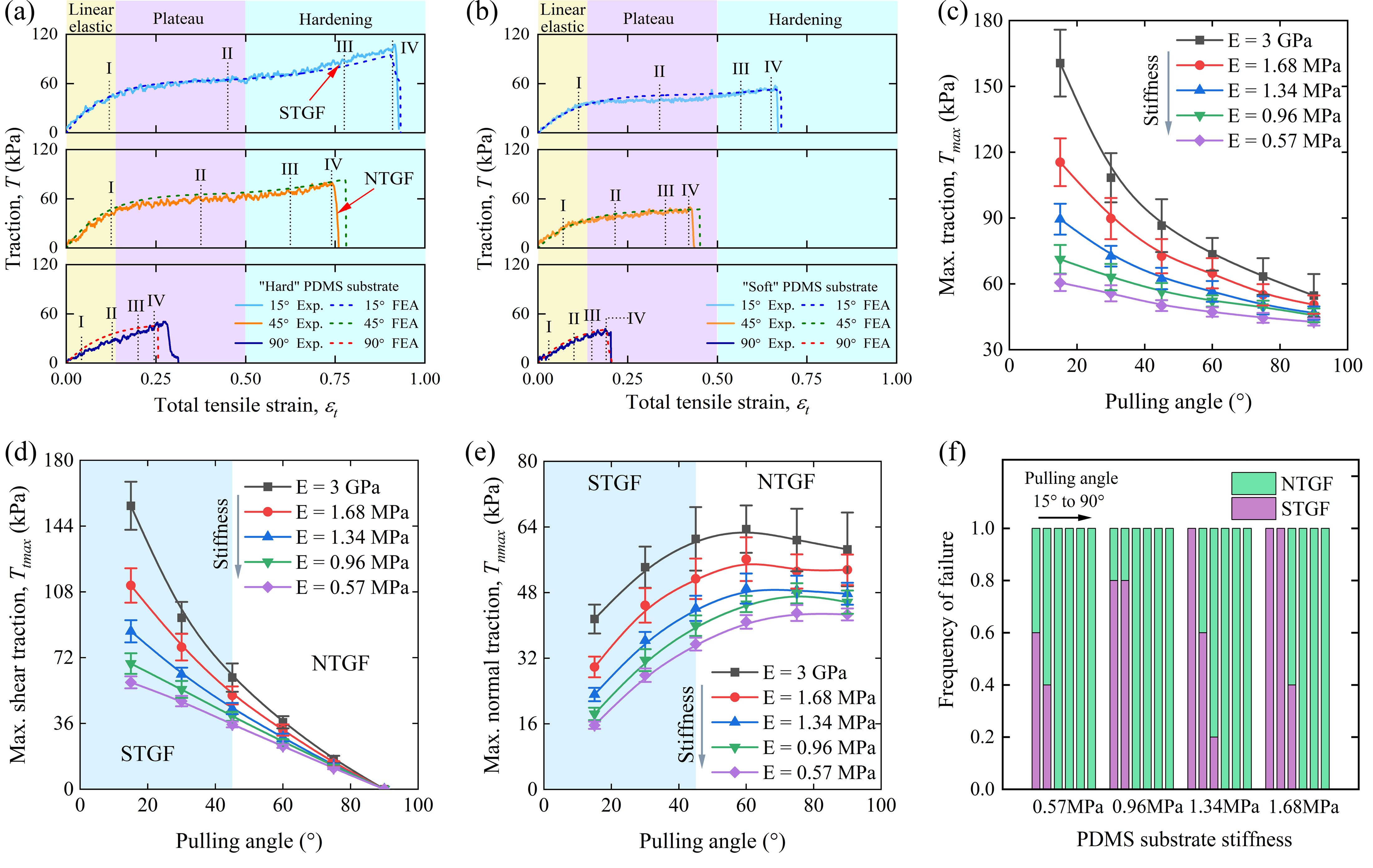}
\caption{\textbf{Mechanical responses of a mussel thread-plaque system under directional tension.} \textbf{a} and \textbf{b} Representative surface traction ($T$) - total strain ($\varepsilon_t$) curves of plaque detachment from “stiff” and “soft” PDMS substrates, respectively. \textbf{c} - \textbf{e} The average maximum traction ($T_{max}$), shear ($T_{smax}$) and normal traction ($T_{nmax}$)  as a function of pulling angle, respectively. Note: the data are means ± SD (n = 5 for each pulling angle) and individual surface traction ($T$) - total strain ($\varepsilon_t$) curves are given in Supplementary Figs. \textcolor{blue}{S5-S9}. \textbf{f} The frequency of STGF and NTGF in response to different combinations of pulling angle and substrate stiffness.}
\label{Fig.3}
\end{figure*}

In a natural environment, the mussel thread-plaque system and the underlying substrate often maintain an angle of less than 45$^\circ$ \cite{Lin2007,Smeathers1979}; however, the pulling angles may exceed this range (i.e., $\theta = 0^\circ - 90^\circ$) under the conditions of strong tidal waves. Therefore, directional tensions were conducted at the pulling angles selected from 0$^\circ$ to 90$^\circ$ to examine the mechanical behaviours of the plaque attachment. In the tests, the thread of all samples was cut in a unified length of approximately 6 mm, which is around 50 times larger than the plaque height of 120 µm. This thread length to plaque height ratio ensured that the measured mechanical responses could represent the detachment behaviour of a real mussel thread-plaque system. Figures \ref{Fig.3}a and b show the representative traction $T -$ total strain $\varepsilon_t$ responses of mussel plaque detached from “stiff” and “soft” PDMS substrates, respectively. For both cases, the value of $\varepsilon_t$ at failure decreases as the pulling angle increases. At a smaller pulling angle (e.g., $\theta = 15^\circ$), the mechanical response of a plaque attachment can be characterised by four regimes: linear elastic, plateau, hardening, and failure. However, as the pulling angle increases, a plaque may fail before reaching the hardening regime.

To investigate the effects of substrate stiffness and pulling angle on the mechanical responses, the average maximum traction $T_{max}=max\left\{T\right\}$ as a function of the pulling angle $\theta$ was plotted in Fig. \ref{Fig.3}c for selected substrate stiffness. Consistent with the observations of existing research \cite{Desmond2015}, under the identical substrate stiffness condition, the value of $T_{max}$ decreases as the pulling angle increase; the stiffness of the substrate has significant impact on the value of $T_{max}$ at a smaller pulling angle, say $\theta = 15^\circ$, i.e., the value of $T_{max}$ increases as the stiffness of the substrate does. However, the effect of the substrate stiffness diminishes as the pulling angle increases: the effect of the substrate stiffness becomes insignificant at $\theta = 90^\circ$. This is a striking phenomenon that has not been reported before. To examine the failure mechanisms, the maximum shear and normal components, $T_{tmax}=max\left\{T_t\right\}$ and $T_{nmax}=max\left\{T_n\right\}$, of the surface traction $T$  as a function of the pulling angle $\theta$ are shown in Figs. \ref{Fig.3}d and e, respectively. The maximum shear component $T_{tmax}$ is sensitive to the pulling angle: the values at $\theta \geq 45^\circ$ are less than 50\% of that at $\theta =15^\circ$. As the plaque detachment was primarily caused by the normal component $T_n$ at $\theta =90^\circ$, one can estimate the strength of the adhesion under normal surface traction based on the response at $\theta =90^\circ$ (Fig. \ref{Fig.3}e). In contrast, the maximum normal traction $T_{nmax}$ is less sensitive to the pulling angle: the values at $\theta \geq 45^\circ$ remain approximately similar to the strength of the adhesion. These results suggest that (1) the plaque adhesion may fail with either normal traction-governed failure (NTGF) mode when pulling angles were greater than $45^\circ$, or shear traction-governed failure (STGF) mode when pulling angles were less than $45^\circ$; (2) NTGF is less sensitive to the substrate stiffness in comparison with STGF; (3) STGF is caused by higher surface traction in comparison with that causing NTGF; and (4) plaques failed with STGF mode have higher load bearing capacity than those failed with NTGF mode.

\begin{figure*}[!tp]
\centering
\includegraphics[width=0.95 \linewidth]{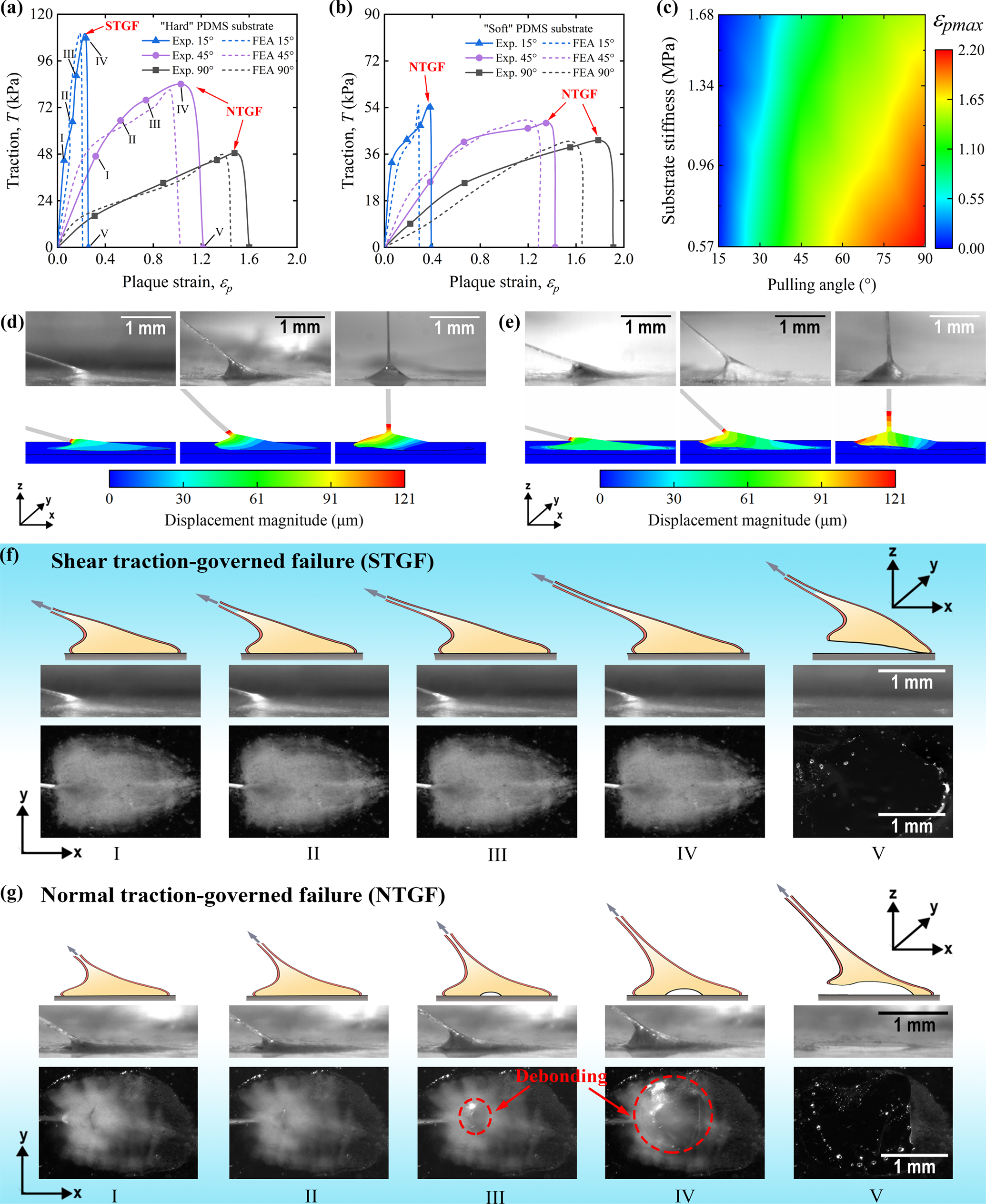}
\caption{\textbf{Deformation of mussel plaques under directional tension.} \textbf{a} and \textbf{b} A comparison of the FE-predicted strain of the mussel plaque with experimental measurements at loading stages I to IV, detaching mussel plaques from “stiff” and "soft" substrates, respectively. \textbf{c} The maximum strain of the mussel plaque ($\varepsilon_{p max}$) in response to different substrate stiffness and pulling angle. $\varepsilon_{p max}$ are the mean value of five individual measurements from the experiments. \textbf{d} and \textbf{e} The side view of FE predicted maximum plaque deformation against experimental images, detaching mussel plaques from “stiff” and "soft" substrates, respectively. \textbf{f} and \textbf{g} Schematic side views in conjunction with time-lapse plaque projections to show the evolution of STGF and NTGF from I to V of Figs. \ref{Fig.4}a and b, respectively.}
\label{Fig.4}
\end{figure*}

To question if pulling angle is the primary factors that determined the failure mode, a series of experimental measurements were conducted to measure the frequency of occurrence for different failure modes, as shown in Fig. \ref{Fig.3}f . When the pulling angle exceeded $45^\circ$, all mussel plaques failed in the mode of NTGF. The high-level determinacy observed may indicate that the pulling angle was the primary contributing factor under these conditions. However, when the pulling angle was less than $45^\circ$, which was closer to the natural loading scenarios, the frequency of STGF mode varied from 0.4 to 1 under the substrate stiffness varied from 0.57 MPa to 1.68 MPa. The level of uncertainty observed may indicate that, under smaller pulling angles, the failure mode may also be influenced by other factors, such as the stiffness of the substrate. 

\subsection*{Responses of a mussel plaque under directional tensions}
The two distinct failure modes, i.e., NTGF and STGF, are associated with two different types of deformation patterns of mussel plaques. Figures \ref{Fig.4}a and b show the FE predicted and measured functional relations between applied surface traction $T$ and the plaque strain $\varepsilon_p$ for selected loading directions for the mussel plaques detached from the “stiff” and the “soft” PDMS substrates, respectively. The FE predictions agree well with experimental measurement. The failure mode for each scenario has been highlighted in these figures, which suggests that, for both failure modes, the surface traction $T$ increased monotonically with plaque strain  until the final catastrophic failure occurred. The FE predictions and experimental measurements suggest that (1) the plaque detached with STGF mode exhibited a much smaller value of plaque strain $\varepsilon_p$ at failure than those detached with the NTGF mode (Fig. \ref{Fig.4}a); (2) an increase in substrate stiffness from 0.57 to 1.68 MPa resulted in a moderate reduction in plaque strain at failure (Fig. \ref{Fig.4}c);  and (3) plaque strain $\varepsilon_p$ at failure increased as the pulling angle did while total strain ($\varepsilon_t$) experienced an opposite trend (Figs. \ref{Fig.4}a and b). 

To further reveal the deformation mechanism, Figs. \ref{Fig.4}f and g show the montages of selected bottom views and side views during the loading history of the plaques that failed with NTGF (i.e., pulling angle $\theta=15^\circ$) and STGF modes ($\theta=45^\circ$), respectively, as indicated in Fig. \ref{Fig.4}a. To aid in the interpretation of the results, the figures also include schematic side views depicting the debonding processes. Under the STGF mode, the mussel plaque exhibited a sudden, catastrophic manner: no interfacial damage was observed until immediate before failure  (Point IV). However, under the NTGF mode, the plaque failed in a progressive manner: debonding was initiated underneath the location of the thread (Point II) and propagated rapidly until fully deboned from the substrate (Point V). It is noted that, although the NTGF mode exhibits progressive failure mode at the interface, there is an increase in surface traction $T$ over the entire process of debonding (Point I to Point IV), as shown in Figs. \ref{Fig.4}a and b. This is distinct from conventional engineering systems, which typically exhibit a noticeable and progressive reduction in load-bearing capacity as they undergo progressive failure. 

\begin{figure*}[!b]
\centering
\includegraphics[width=0.85 \linewidth]{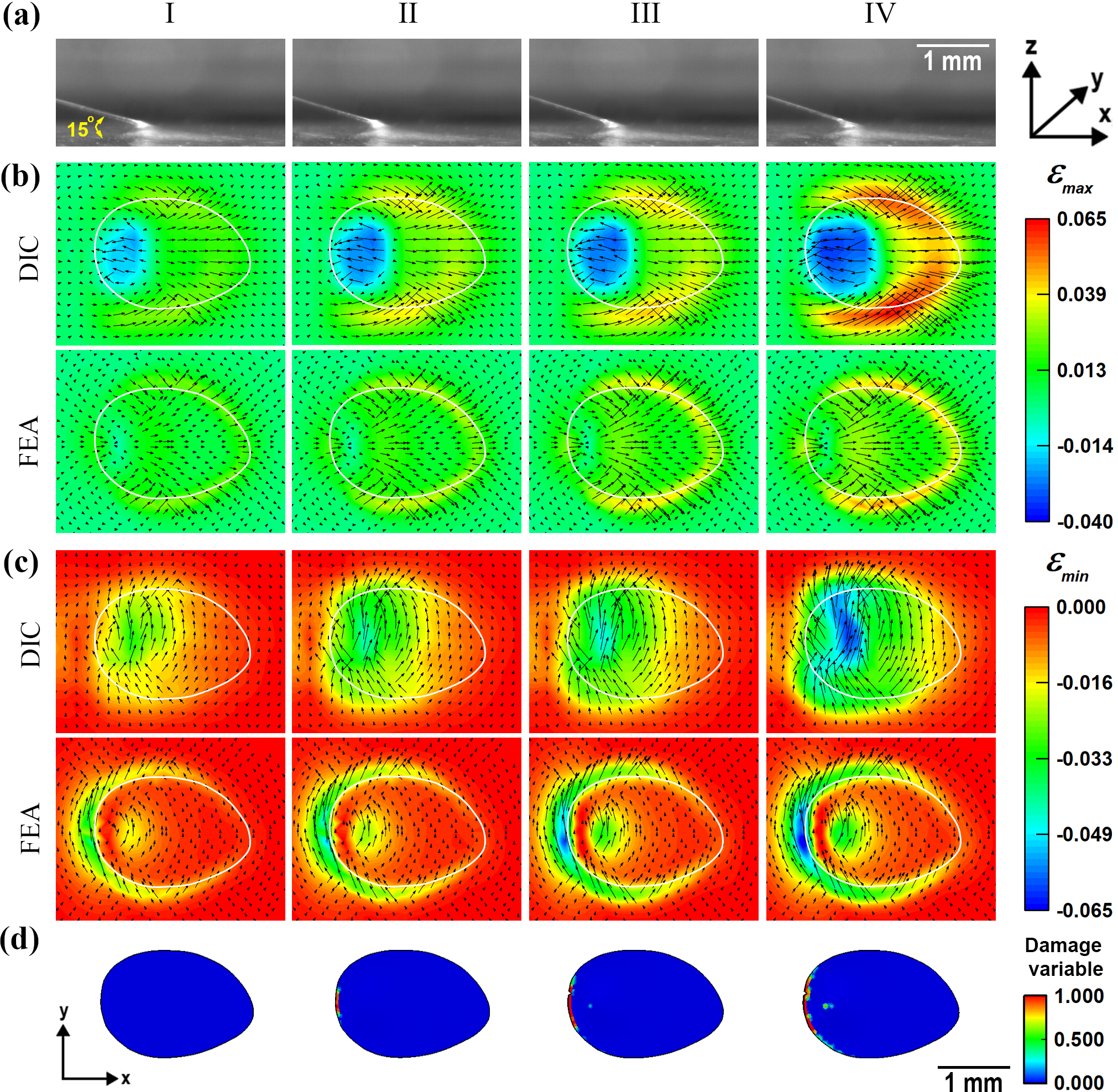}
\caption{\textbf{Mussel plaque and substrate interaction on “hard” substrate under $15^\circ$ tension.} \textbf{a} The side view of mussel plaque detaching under $15^\circ$ tension. \textbf{b} and \textbf{c} Comparisons of in-situ DIC measurements and FE predictions showing the evolution of $\varepsilon_{max}$ and $\varepsilon_{min}$ on “hard” substrate, respectively. \textbf{d} The evolution of FE predicted cohesive failure on “hard” substrate. Note that there is no debonding failure at the interface, and the scalar damage variables are below the threshold of element deletion. The contour in white colour indicates the edge of real mussel plaques.}
\label{Fig.5}
\end{figure*}

\subsection*{In-situ DIC measurement on plaque and substrate interactions}
The deformation patterns exhibited by a deformable substrate under a plaque-substrate interaction event can provide clues on how the traction forces are transmitted from a plaque to the underlying substrate, which can further strengthen our understanding on the interaction event. We employed in-situ Digital Image Correlation (DIC) to measure the in-plane maximum and minimum principal strains, $\varepsilon_{max}$ and $\varepsilon_{min}$, on the top surface of the “stiff” and “soft” substrates during directional tensile tests. To ensure that the mfps at the interfaces were fully developed, the plaques were brought into contact with the underlying substrates for a period of 48 hours prior to the tensile tests. FE predictions were also conducted for the purposes of interpretation and comparison.

\begin{figure*}[!b]
\centering
\includegraphics[width=0.85 \linewidth]{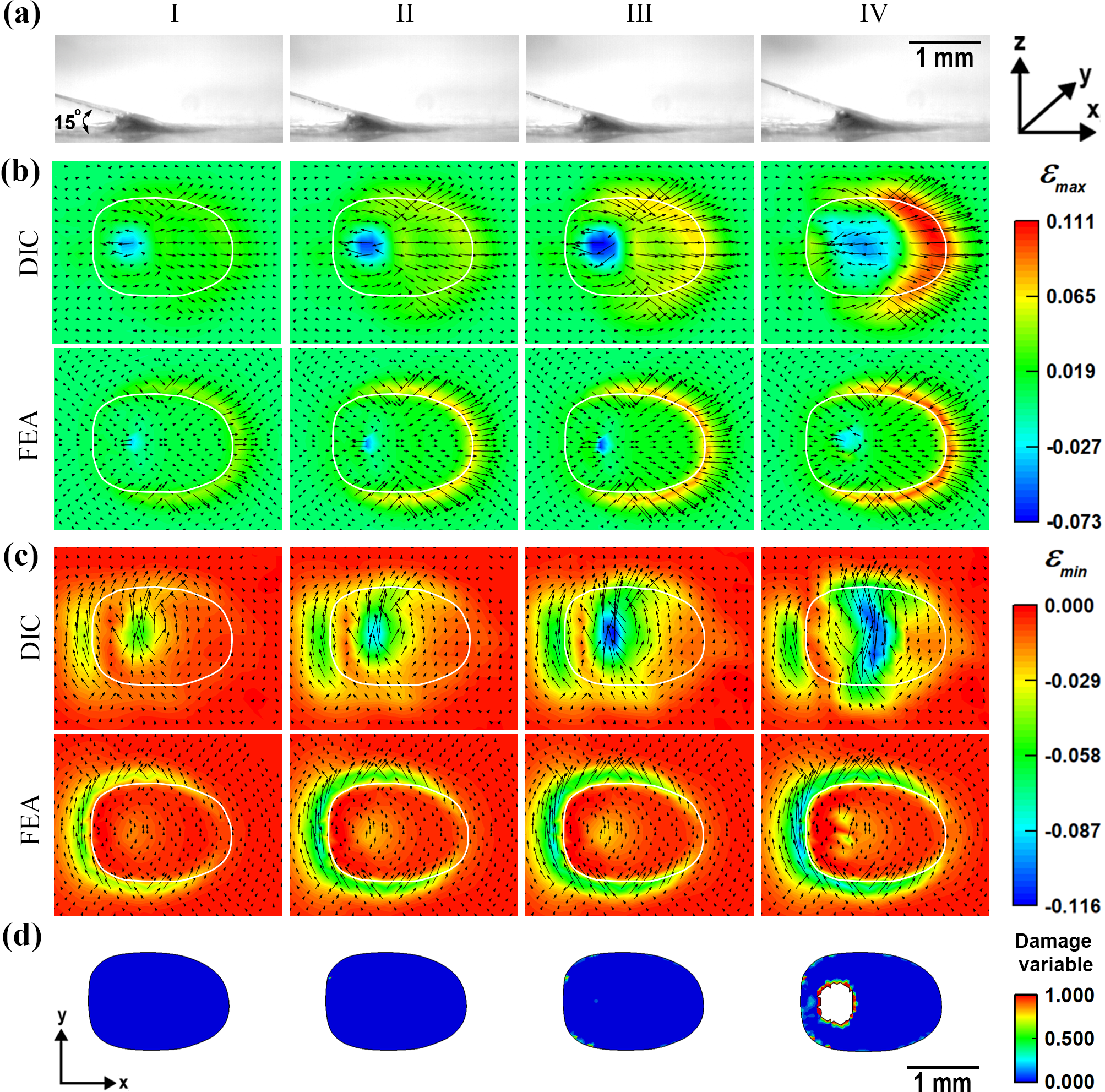}
\caption{\textbf{Mussel plaque and substrate interaction on “soft” substrate under $15^\circ$ tension.} \textbf{a} The side view of mussel plaque detaching under $15^\circ$ tension. \textbf{b} and \textbf{c} Comparisons of DIC measurements and FE predictions showing the evolution of $\varepsilon_{max}$ and $\varepsilon_{min}$ on “soft” substrate, respectively. \textbf{d} The evolution of FE predicted cohesive failure on a “soft” substrate. Note that debonding failure occurred at Point IV and the scalar damage variables exceeded the threshold of element deletion. The contour in white colour indicates the edge of real mussel plaques.}
\label{Fig.6}
\end{figure*}

Figure \ref{Fig.5} presents the montages at selected loading stages, showing the side view of plaque deformation, in-plane principal strains and FE-predicted damage variable at the interface (cohesive element) for the plaque that was detached from the “stiff” substrate under $15^\circ$ tension and failed under the STGF mode. In-situ DIC and FE simulation show that the vicinity directly beneath the intersection of the thread and plaque, i.e. the rear portion, experienced compression in both in-plane principal directions. Moving away from the rear portion towards the front portion, the maximum principal strain $\varepsilon_{max}$ of the underlying substrate gradually shifted from compression to tension (Fig. \ref{Fig.5}b), while the minimum principal strain $\varepsilon_{min}$ remained in compression (Fig. \ref{Fig.5}c). The highest tensile strain was observed around the front and side boundaries of the mussel plaque, with principal angles varying from $-43.94^\circ$ to $44.14^\circ$ with respect to the outward normal of the plaque boundary (Fig. \ref{Fig.5}b). The FE predictions show a reasonable agreement with the in-situ DIC measurements. Consistent with experimental observation (Fig. \ref{Fig.4}f), FE predictions demonstrated that no significant debonding occurred within the interaction prior to failure, and the interface failed in a sudden and catastrophic manner (Fig. \ref{Fig.5}d). During the loading process, the mussel plaque experienced insignificant deformation (Fig. \ref{Fig.5}a), and the geometrical sizes and positions of the compression and tension zones (Figs. \ref{Fig.5}b and c) remained fairly constant due to the interface's lack of debonding.

\begin{figure*}[!b]
\centering
\includegraphics[width=0.85 \linewidth]{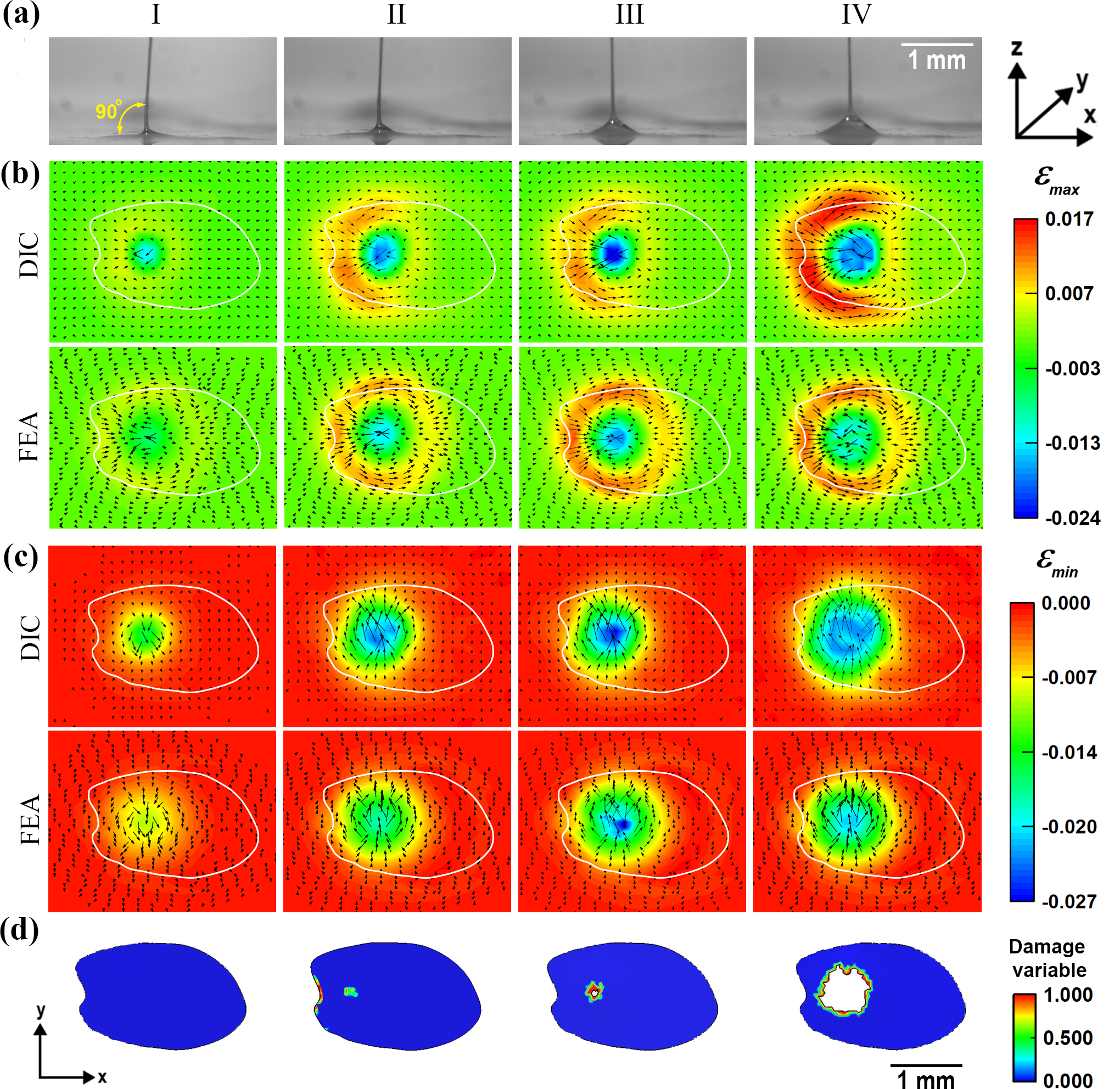}
\caption{\textbf{Mussel plaque and substrate interaction in response to $90^\circ$.} \textbf{a} The side view of mussel plaque detaching from the substrate under 90º tension. \textbf{b} and \textbf{c} Comparisons of DIC measurements and FE predictions showing the evolution of $\varepsilon_{max}$ and $\varepsilon_{min}$ under $90^\circ$ tension, respectively. \textbf{d} The evolution of FE predicted cohesive failure under $90^\circ$ tension. The contour in white colour indicates the edge of real mussel plaques. Note that “stiff” substrates are used here to compare FE simulation and in-situ DIC measurement. The deformation pattern at the “soft” PDMS substrate was given in Supplementary Fig. \textcolor{blue}{S12}.}
\label{Fig.7}
\end{figure*}

The plaques detached from the “soft” substrate under $15^\circ$ tension failed under the NTGF mode (Fig. \ref{Fig.4}b). During the loading process, the mussel plaque exhibited a noticeable deformation, as shown in Point IV of Fig. \ref{Fig.6}a. Again, the highest in-plane tensile strain in the substrate was observed around the front and side boundaries of the mussel plaque (Fig. \ref{Fig.6}b), and the substrate around the rear portion experienced compression in both in-plane principal directions (Figs. \ref{Fig.6}b and c). The interfaces failed in a progressive failure mode, with debonding starting from the rear portion and gradually spreading towards the front portion (Fig. \ref{Fig.6}d). As a result, the geometrical sizes and positions of the compression and tension zones constantly change, with the compression zone expanding and shifting towards the front boundary (Fig. \ref{Fig.6}b). The plaques under $45^\circ$ tension, attached on both the “soft” and “stiff” substrates, exhibit a similar deformation pattern, as shown in Supplementary Figs. \textcolor{blue}{S10-S11}).

The deformation pattern of the “stiff” substrate caused by the $90^\circ$ tension is distinct from those caused by $15^\circ$ or $45^\circ$ tensions, as shown in Figs. \ref{Fig.7}b and c. The interfaces failed in a progressive failure mode, with debonding starting from the centre of the rear portion and gradually extending towards the outer edge of the rear portion (Fig. \ref{Fig.7}d). Both the FE prediction and the in-situ DIC measurement show that the distribution of the maximum in-plane strain $\varepsilon_{max}$ exhibits a ring pattern with compression in the interior and tension in the exterior (Fig. \ref{Fig.7}b). The distribution of the minimum in-plane strain $\varepsilon_{min}$ exhibits a circle pattern with the highest in-plane compression in the centre and decreases radially (Fig. \ref{Fig.7}c). Debonding rapidly developed from III to IV, which caused a reduction of strain intensity at the compression zone (Fig. \ref{Fig.7}d). The mussel plaque deforms significantly owing to the presence of debonding (Fig. \ref{Fig.7}a). The substrate around the rear portion is the only area where the in-plane principal strains were detected. The absence of in-plane strains in the front portion may suggest that the front portion did not contribute to the in-plane loading bearing. The deformation pattern of the “soft” substrate under $90^\circ$ tension is similar to that of the “stiff” substrate (Supplementary Fig. \textcolor{blue}{S12}).

\section*{Discussions}
Marine mussels have evolved a plaque-thread system to withstand the significant forces imposed by turbulent tidal currents and predator attacks. Recent research progress has suggested that the porous structure in plaque core is reminiscent of cellular solids \cite{Chisca2018,Yang2022,Shi2021} or lattice structures \cite{Pham2019,Liu2021,Shaikeea2022,Bhuwal2023,Yang20222} with enhanced load-bearing and damage-tolerant properties. Nevertheless, prior studies have not elucidated the underwater adhesive mechanisms of mussel plaque \cite{Desmond2015,Filippidi2015,Bernstein2020,Priemel2017}. In this groundbreaking study, we present the first investigation into plaque/substrate interaction and provide insights into the deformation patterns occurring on various substrates.

The present study investigated the failure mechanisms of underwater plaque detachment. Consistent with previous investigations \cite{Desmond2015,Waite2017}, NTGF is a prevalent failure mode that was observed when mussel plaques were detached from substrates at a pulling angle greater than 45°. However, a distinct  STGF was found for the first time when mussel plaques were detached from "stiff" substrates at a small pulling angle (15° tension) closely aligned with the plaque's natural angle. Notably, this study deviates from earlier research by not finding experimental evidence of interior and exterior cohesive failures. This could be attributed to differences in mussel species or testing environments (moist or dry conditions) employed in prior studies. Furthermore, the previous investigation reported that pulling angles is the primary factor affecting the failure modes of plaque detachment \cite{Desmond2015}. However, the present study emphasises the significance of substrate stiffness and reveals a synergistic effect of pulling angle and substrate stiffness on plaque detachment. Our study found that the plaque-thread system failed in the STGF mode exhibits higher surface traction (Fig. \ref{Fig.3}c) and less plaque deformation than the NTGF (Figs. \ref{Fig.4}a-c), which demonstrates the mechanical advantages of anchoring to "stiff" substrates at a small pulling angle. 

Deformation patterns at substrates were investigated using 2D in-situ DIC measurements in conjunction with FE simulation. A method of fabricating PDMS substrate (Fig. \ref{Fig.2}a) for DIC measurements was developed to track the substrate deformation under plaque tension. The findings indicated that under 15º tension, plaque contraction could introduce in-plane tension and compression at the front and rear portion of the plaque, respectively. As the pulling angle increase from 15º to 90º, the tension zone in $\varepsilon_{max}$ shifted towards the rear portion of the plaque and exhibited a "ring" pattern with interior compression and exterior tension (Fig. \ref{Fig.7}b), indicating that the front portion of the plaque did not contribute to the in-plane loading bearing . A reduction in strain intensity was observed in the compression zone of $\varepsilon_{max}$, which is a typical feature of NTGF mode, whereas its intensity increased monotonically as the tensile load in STGF mode. Additionally, our study found that increasing substrate stiffness or reducing the pulling angle can prevent or delay the onset of NTGF, resulting in a failure mode change from NTGF to STGF and ultimately enhancing the surface traction of the plaque-thread system.

The mechanical lessons learned from plaque/substrate interaction could open several new research directions. In particular, this work inspires general strategies or design principles to develop interfacial adhesives for joining materials with distinct mechanical properties. Also, it sheds light on the development of plaque-inspired adhesive structures for artificial deep-sea anchoring systems of floating wind turbines. Further studies should focus on optimising the current FE model and experimental methods. These could include measuring the mechanical property and concentration of mfp at the interface, developing a new cohesive law at the molecule level, creating a real plaque adhesive structure based on Nano CT scanning or confocal laser scanning, and investigating the traction force distribution at the substrates.

\section*{Methods}
\subsection*{Mussel samples preparation}
Blue mussels (\emph{Mytilus edulis}) were collected from the Hunstanton mussel farm ($52.94^\circ$N, $0.49 ^\circ$E) in England (Supplementary Fig. \textcolor{blue}{S1}a) and kept in an insulated cooler with natural seawater (temperature = $7^\circ$C, and salinity = 33 ppt) during the shipment. Before placing mussel samples in the aquarium, the salinity tester was calibrated by a 35.00 ppt calibration standard sachet, and then the salinity and temperature of the artificial seawater were checked by a Hanna HI-98319 tester. These mussels were further placed in a laboratory aquarium ($600 \times1200\times600$ mm) filled with continuously circulated seawater (Supplementary Fig. \textcolor{blue}{S1}b). Sea salt (Tropic Marin Pro-Reef Sea Salt, UK) was dissolved in fresh water to simulate natural seawater conditions (salinity = 33 ppt and pH = 8). An aquarium chiller was employed to maintain temperatures of the seawater ranging from $5^\circ$C to $7^\circ$C, and an aeration system was used to supply adequate oxygen into the artificial seawater. A daily, partial water change, replacing 20\% of the total water volume was implemented to maintain the water quality. Shellfish Diet 1800 (Reed Mariculture, USA), a mixture of marine microalgaes, was employed to feed the mussels daily. Mussel samples were fastened with substrates (Supplementary Fig. \textcolor{blue}{S1}c) to allow the deposition of plaque for 48 hours, and their threads were carefully cut for SEM imaging and mechanical tests.

\subsection*{Scanning electron microscopy (SEM)}
To prepare cryo-sections for SEM imaging, plaque samples were fixed in a solution of 3.7\% formaldehyde and 2.5\% glutaraldehyde for approximately four hours and then placed into Milli-Q water for three days at $0^\circ$C to $5^\circ$C. These samples were quickly frozen in the embedding medium (Thermo Scientific OCT) at $-40^\circ$C and sliced into $40\mu m$ thick cryo-sections at a cryo-chamber temperature of $-20^\circ$C using Leica cryostat (CM3050S). Cryo-sections were further rinsed three times in Milli-Q water to remove the cryoprotectant. An incremental solvent exchange method from Milli-Q water to ethanol to hexamethyldisilazane (HMDS) was employed to avoid damage to plaque samples during the freeze-drying process \cite{Nordestgaard1985,Bray1993,Filippidi2015}. The mass ratio of ethanol in water/ethanol exchange was altered from 0\% to 100\% at an increment of 25\%, and then the mass ratio of HMDS in ethanol/HMDS exchange underwent an increase from 0\% to 100\% at an increment of 33\%. Finally, SEM samples were sputter coated with 10 nm thick iridium layers for two minutes and imaged in JEOL 7000 SEM using an accelerating voltage of 5 $kV$.

\subsection*{Substrate fabrication}
PDMS was employed to fabricate deformable substrates. Dow Corning Sylgard 184 silicone elastomer (10:1 wt.\% for base and curing agent) and Sylgard 527 silicone gel (1:1 wt.\% for parts A and B) were mixed with different mass ratios to tune Young's modulus from $0.57 \pm  0.01$ MPa to $1.68 \pm  0.05$ MPa (Supplementary Table \textcolor{blue}{S1}). The mechanical properties of different PDMS formulations (Supplementary Figs. \textcolor{blue}{S2}b-e) were characterised using standard tensile specimens according to ASTM D412 \cite{ASTMD412} (Supplementary Fig. \textcolor{blue}{S2}a). Once the stiffness of different PDMS formulations was determined, 4-layer substrates (Fig. \ref{Fig.2}a) were built using spin coating with different combinations of spin speed and duration (Supplementary Table \textcolor{blue}{S2}). Microparticles (ZnS:Cu \cite{Moon2013,Park2019}) were mixed with PDMS (5:1 wt.\% for PDMS and Micro-particles) to create the particle layer in Fig. \ref{Fig.2}a. Black liquid silicone pigment (Easycomposites, UK) was added into the PDMS at a mass ratio of 1\% to fabricate the particle layer in Fig. \ref{Fig.2}a. Since the surface wettability and roughness could affect the underwater adhesion \cite{Stark2013}, water contact angle (WCA) and atomic force microscopy (AFM) was employed to confirm consistency in the surface wettability and topography among different PDMS substrates, respectively. Representative WCAs and surface topographies exhibit insignificant differences between different PDMS substrates (Supplementary Figs. \textcolor{blue}{S3}c and d). Thus, the surface wettability and roughness can be treated as the control variables in this study. 

\subsection*{Mechanical tests}
A customised microscopy system was developed for mechanical tests (Supplementary Figs. \textcolor{blue}{S4}a and b). Directional tensions were applied to mussel threads via a linear actuator (Thomson MLA11A05), at a loading speed of 50 $\mu m/s$. This loading speed was sufficiently slow to measure the quasi-static responses of the plaque-thread systems as the measured responses were not sensitive to further reduction of the loading speed. A high-resolution load cell (a precision of 0.01N) (Honeywell Model 34) was attached between the rigid clamp and linear actuator to measure the load response during the tests. Displacement and force data were recorded by a custom-written program using an Arduino Nano board. Randomly distributed microparticles provided a strong reflection under the illumination of a LED light (Edmund Optics, 530 nm). The $12\times$ zoom imaging systems equipped with two CCD cameras (Pixelink PL-D USB 3.0 CMOS) were employed to capture deformations of plaques and substrates. In the tests, the CCD camera was configured to match the loading speed at a frame rate of 50 fps.

\subsection*{Two dimensional (2D) in-situ digital image correlation (DIC)}
2D in-situ DIC measurement was implemented in DICe software \cite{DICe2015}, and microparticles (ZnS:Cu) \cite{Moon2013,Park2019}, as tracking features for full-field displacement measurements, were randomly distributed in PDMS (Supplementary Fig. \textcolor{blue}{S3}a). The particle size followed a normal distribution in all substrates, and the particle radius (mean ± standard deviation) was between 10.26 ± 5.32 µm and 10.80 ± 5.36 µm (Supplementary Fig. \textcolor{blue}{S3}b). The pattern of microparticles was imaged using the microscopy system at a spatial resolution of 36 pixels/mm. The region of interest in DIC measurement was 600 × 600 pixels, and the step size was configured as 32 pixels. The subset image was set as 32 × 32 pixels to include approximately 6 to 9 particles, which ensures the optimal particle density of around 25\% to 50\% for DIC measurement \cite{Phillip2015,Pang2020,Su2019}.

\subsection*{Finite element (FE) simulations}
Each plaque-thread system was modeled by five components: a thread, a cuticle layer, a plaque core, a cohesive interface, and a substrate, as shown in Figs. \ref{Fig.2}f and i.
Since the thread and the cuticle layer exhibit a similarly microstructure, they were modelled as an identical material using
three-order Ogden's model \cite{Ogden1972}
\begin{equation}
\begin{split}
\Psi (\bar{\lambda}_{1},\bar{\lambda}_{2},\bar{\lambda}_{3})=\sum_{k=1}^{N}\frac{2\mu _{k}}{\alpha _{k}^{2}}(\bar{\lambda}_{1}^{\alpha_{k}}+\bar{\lambda}_{2}^{\alpha_{k}}+\bar{\lambda}_{3}^{\alpha_{k}}-3)
\end{split}
\label{Eq7}
\end{equation}
where $\bar{\lambda}_{k}$ are distortional principal stretches which can be calculated from principal stretches ($\lambda_{k}$) by $\bar{\lambda}_{k}=J^{-1/3}\lambda_{k}$, $N$ is the order of the strain energy potential ($N = 3$), $k$ is the index of summation ($k =1, 2, 3$), $J$ is the elastic volume strain, $J={\lambda}_1 {\lambda}_2 {\lambda}_3$, and $\mu_k$ and $\alpha_k$ are the material constants. As shown in Fig. \ref{Fig.2}k, the mechanical response of the mussel thread can be captured by Ogden’s model using the material data shown in Table \ref{Tab.1}.

The mfps at the plaque and substrate interface were modelled with 6-node three-dimensional cohesive elements (COH3D6 elements). Let $t_s$ denotes the $y$-direction tangential component of the local surface traction, and $\delta_n$, $\delta_t$ and $\delta_s$ are the normal and shear separations corresponding to tractions. The adhesive behaviour of mfps was defined by the following traction versus separation relation \cite{Camanho2002}
\begin{equation}
t_n=\left\{\begin{matrix}
(1-D)k_n\delta_n \textrm{, for }\delta_n> 0\\ 
\\ 
k_n\delta_n \textrm{, otherwise}
\end{matrix}\right.
\label{Eq8}
\end{equation}
\begin{equation}
t_s=(1-D)k_s\delta_s
\label{Eq9}
\end{equation}
\begin{equation}
t_t=(1-D)k_t\delta_t
\label{Eq10}
\end{equation}
where $k_n$ and $k_s$ are the normal and shear stiffness, respectively, and the damage variable $D\in [0,1]$ with the initial value of 0. For compression, i.e., $\delta_n \leq 0$, the damage variable does not affect the cohesive relation. The onset of damage was assumed when the following quadratic interaction criterion reaches a value of unity  
\begin{equation}
\left \{ \frac{t_{n}}{t_{n}^{o}} \right\}^2+\left \{ \frac{t_s}{t_{s}^{o}} \right\}^2+\left \{ \frac{t_{t}}{t_{t}^{o}} \right\}^2=1
\label{Eq11}
\end{equation}
where $t_n^o$, $t_s^o$ and $t_t^o$ denote the maximum traction when the separation is either purely normal or pure shear, respectively. To define the damage evolution of the cohesive layer, linear softening law based on fracture energy was employed. Define an effective separation by 
\begin{equation}
\delta_e=\sqrt{\left\langle\delta_n\right\rangle^2+\delta_s^2 + \delta_t^2}
\label{Eq12}
\end{equation}

The work conjugated effective traction to this effective separation can be defined as
\begin{equation}
t_e=\sqrt{\left\langle t_n\right\rangle^2+t_s^2 + t_t^2}
\label{Eq13}
\end{equation}

The damage variable $D$ is then defined in terms of the fracture energy $G_c$ as 
\begin{equation}
D=\frac{\frac{2G_c}{t_e^o}(\delta_{e}^{max}-\delta_{e}^{o})}{\delta_{e}^{max}(\frac{2G_c}{t_e^o}-\delta_{e}^{o})}\leq 1
\label{Eq14}
\end{equation}
where $\delta_e^{max}$ is the maximum value of $\delta_e$ attained during the loading history while $t_e^0$ and $\delta_e^0$ are the values of effective traction at the initiation of damage. 

The material parameters that define the interface model are the maximum tractions $t_n^o$, $t_s^o$ and $t_t^o$; the stiffnesses $k_n$, $k_s$ and $k_t$; and fracture energy $G_c$. It was found that soft substrate weakens mussel wet adhesion through mechanosensing \cite{Choi2021,Amini2017,Harrington2018}. Therefore, two sets of interface properties were employed to simulate plaques anchoring to the “stiff” and the “soft” PDMS substrates, respectively, as shown in Table \ref{Tab.1}. The maximum tractions ($t_n^o$, $t_s^o$, and $t_t^o$) and the stiffnesses ($k_n$, $k_s$, and $k_t$) were estimated by calibrating the FE-predicted traction $T$ - total strain $\varepsilon_t$ responses against their experimental results (Figs. \ref{Fig.3}a and b).
Numerical experiments were conducted to ensure that the outcomes of the numerical simulations were independent of the mesh density in FE simulations. Thus, the thread, cuticle layer, plaque core, cohesive interface, and substrate for each FE model were meshed using global sizing control with an approximate global mesh size of 50 µm. To ensure the simulations capture the quasi-static responses of the system, the loading rate was controlled to ensure that the kinetic energy of the system was within 5\% of the total energy.

\begin{table*}[!t]
\renewcommand\tabcolsep{3pt}
\renewcommand{\arraystretch}{1.6}
\centering\setlength{\abovecaptionskip}{5pt}
\caption{Material properties of plaque/substrate system used in FE simulation}
\label{Tab.1}
\begin{tabular}{rccccccc}
    \hlineB{2.5}
    Materials& \multicolumn{7}{c}{Material constants used for FE simulation}\\
    \hlineB{2.5}
    &$\mu _{1}$ ($MPa$)&$\alpha_{1}$ &$\mu _{2}$ ($MPa$)&$\alpha_{2}$ & $\mu _{3}$ ($MPa$) &$\alpha_{3}$\\
                          Thread \& cuticle layer&-134.25&1.29&19.84&2.77&181.07&-3.93\\
                          Plaque core  &-17.08&5.99&8.62&6.64&33.69&-8.21\\
    \hlineB{1}
    &$t_{n}^{o}$&$t_{s}^{o}$&$t_{t}^{o}$&$K_{n}$&$K_{s}$&$K_{t}$&$G_{c}$\\
    &($MPa$)&($MPa$)&($MPa$)&($MPa/mm$)&($MPa/mm$)&($MPa/mm$)&($J/m^2$)\\
                          Cohesive layer on "stiff" PDMS &0.38&0.30&0.30&13000&1200&1200&102\\
                          Cohesive layer on "soft" PDMS &0.16&0.25&0.25&12000&1100&1100&102\\
    \hlineB{1}
    &\multicolumn{3}{c}{Young’s modulus, $E$ (Mpa)}&&\multicolumn{3}{c}{Poisson’s ratio, $v$}\\
                          "stiff" PDMS substrate &\multicolumn{3}{c}{1.68}&&\multicolumn{3}{c}{0.49}\\
                          "soft" PDMS substrate &\multicolumn{3}{c}{0.56}&&\multicolumn{3}{c}{0.49}\\
    \hlineB{2.5}
  \end{tabular}
\end{table*}

\bibliography{Manuscript}

\begin{thebibliography}{10}
\urlstyle{rm}
\expandafter\ifx\csname url\endcsname\relax
  \def\url#1{\texttt{#1}}\fi
\expandafter\ifx\csname urlprefix\endcsname\relax\def\urlprefix{URL }\fi
\expandafter\ifx\csname doiprefix\endcsname\relax\def\doiprefix{DOI: }\fi
\providecommand{\bibinfo}[2]{#2}
\providecommand{\eprint}[2][]{\url{#2}}

\bibitem{Fan2021}
\bibinfo{author}{Fan, H.} \& \bibinfo{author}{Gong, J.~P.}
\newblock \bibinfo{journal}{\bibinfo{title}{Bioinspired underwater adhesives}}.
\newblock {\emph{\JournalTitle{Adv. Mater.}}} \textbf{\bibinfo{volume}{33}},
  \bibinfo{pages}{2102983} (\bibinfo{year}{2021}).

\bibitem{Stewart2011}
\bibinfo{author}{Stewart, R.~J.}, \bibinfo{author}{Ransom, T.~C.} \&
  \bibinfo{author}{Hlady, V.}
\newblock \bibinfo{journal}{\bibinfo{title}{Natural underwater adhesives}}.
\newblock {\emph{\JournalTitle{J. Polym. Sci. B: Polym. Phys.}}}
  \textbf{\bibinfo{volume}{49}}, \bibinfo{pages}{757--771}
  (\bibinfo{year}{2011}).

\bibitem{Chen2020}
\bibinfo{author}{Chen, Y.} \emph{et~al.}
\newblock \bibinfo{journal}{\bibinfo{title}{Bioinspired multiscale wet adhesive
  surfaces: structures and controlled adhesion}}.
\newblock {\emph{\JournalTitle{Adv. Funct. Mater.}}}
  \textbf{\bibinfo{volume}{30}}, \bibinfo{pages}{1905287}
  (\bibinfo{year}{2020}).

\bibitem{Heydari2020}
\bibinfo{author}{Heydari, S.}, \bibinfo{author}{Johnson, A.},
  \bibinfo{author}{Ellers, O.}, \bibinfo{author}{McHenry, M.~J.} \&
  \bibinfo{author}{Kanso, E.}
\newblock \bibinfo{journal}{\bibinfo{title}{Sea star inspired crawling and
  bouncing}}.
\newblock {\emph{\JournalTitle{J. R. Soc. Interface}}}
  \textbf{\bibinfo{volume}{17}}, \bibinfo{pages}{20190700}
  (\bibinfo{year}{2020}).

\bibitem{Green2013}
\bibinfo{author}{Karlsson~Green, K.}, \bibinfo{author}{Kovalev, A.},
  \bibinfo{author}{Svensson, E.~I.} \& \bibinfo{author}{Gorb, S.~N.}
\newblock \bibinfo{journal}{\bibinfo{title}{Male clasping ability, female
  polymorphism and sexual conflict: fine-scale elytral morphology as a sexually
  antagonistic adaptation in female diving beetles}}.
\newblock {\emph{\JournalTitle{J. R. Soc. Interface}}}
  \textbf{\bibinfo{volume}{10}}, \bibinfo{pages}{20130409}
  (\bibinfo{year}{2013}).

\bibitem{Tramacere2013}
\bibinfo{author}{Tramacere, F.} \emph{et~al.}
\newblock \bibinfo{journal}{\bibinfo{title}{The morphology and adhesion
  mechanism of octopus vulgaris suckers}}.
\newblock {\emph{\JournalTitle{PLoS One}}} \textbf{\bibinfo{volume}{8}},
  \bibinfo{pages}{e65074} (\bibinfo{year}{2013}).

\bibitem{Meng2019}
\bibinfo{author}{Meng, F.} \emph{et~al.}
\newblock \bibinfo{journal}{\bibinfo{title}{Tree frog adhesion biomimetics:
  opportunities for the development of new, smart adhesives that adhere under
  wet conditions}}.
\newblock {\emph{\JournalTitle{Philos. Trans. R. Soc. A}}}
  \textbf{\bibinfo{volume}{377}}, \bibinfo{pages}{20190131}
  (\bibinfo{year}{2019}).

\bibitem{Ditsche2019}
\bibinfo{author}{Ditsche, P.} \& \bibinfo{author}{Summers, A.}
\newblock \bibinfo{journal}{\bibinfo{title}{Learning from northern clingfish
  (gobiesox maeandricus): bioinspired suction cups attach to rough surfaces}}.
\newblock {\emph{\JournalTitle{Philos. Trans. R. Soc. B: Biol. Sci.}}}
  \textbf{\bibinfo{volume}{374}}, \bibinfo{pages}{20190204}
  (\bibinfo{year}{2019}).

\bibitem{Lin2009}
\bibinfo{author}{Lin, A.}, \bibinfo{author}{Brunner, R.},
  \bibinfo{author}{Chen, P.}, \bibinfo{author}{Talke, F.} \&
  \bibinfo{author}{Meyers, M.}
\newblock \bibinfo{journal}{\bibinfo{title}{Underwater adhesion of abalone: The
  role of van der waals and capillary forces}}.
\newblock {\emph{\JournalTitle{Acta Mater.}}} \textbf{\bibinfo{volume}{57}},
  \bibinfo{pages}{4178--4185} (\bibinfo{year}{2009}).

\bibitem{David2015}
\bibinfo{author}{Labonte, D.} \& \bibinfo{author}{Federle, W.}
\newblock \bibinfo{journal}{\bibinfo{title}{Scaling and biomechanics of surface
  attachment in climbing animals}}.
\newblock {\emph{\JournalTitle{Philos. Trans. R. Soc. B: Biol. Sci.}}}
  \textbf{\bibinfo{volume}{370}}, \bibinfo{pages}{20140027}
  (\bibinfo{year}{2015}).

\bibitem{David2014}
\bibinfo{author}{Slater, D.~M.}, \bibinfo{author}{Vogel, M.~J.},
  \bibinfo{author}{Macner, A.~M.} \& \bibinfo{author}{Steen, P.~H.}
\newblock \bibinfo{journal}{\bibinfo{title}{Beetle-inspired adhesion by
  capillary-bridge arrays: pull-off detachment}}.
\newblock {\emph{\JournalTitle{J. Adhes. Sci. Technol.}}}
  \textbf{\bibinfo{volume}{28}}, \bibinfo{pages}{273--289}
  (\bibinfo{year}{2014}).

\bibitem{Yang2020}
\bibinfo{author}{Yang, J.}, \bibinfo{author}{Bai, R.}, \bibinfo{author}{Chen,
  B.} \& \bibinfo{author}{Suo, Z.}
\newblock \bibinfo{journal}{\bibinfo{title}{Hydrogel adhesion: a supramolecular
  synergy of chemistry, topology, and mechanics}}.
\newblock {\emph{\JournalTitle{Adv. Funct. Mater.}}}
  \textbf{\bibinfo{volume}{30}}, \bibinfo{pages}{1901693}
  (\bibinfo{year}{2020}).

\bibitem{Park2017}
\bibinfo{author}{Park, H.-H.} \emph{et~al.}
\newblock \bibinfo{journal}{\bibinfo{title}{Flexible and shape-reconfigurable
  hydrogel interlocking adhesives for high adhesion in wet environments based
  on anisotropic swelling of hydrogel microstructures}}.
\newblock {\emph{\JournalTitle{ACS Macro Lett.}}} \textbf{\bibinfo{volume}{6}},
  \bibinfo{pages}{1325--1330} (\bibinfo{year}{2017}).

\bibitem{Amador2017}
\bibinfo{author}{Amador, G.~J.}, \bibinfo{author}{Endlein, T.} \&
  \bibinfo{author}{Sitti, M.}
\newblock \bibinfo{journal}{\bibinfo{title}{Soiled adhesive pads shear clean by
  slipping: a robust self-cleaning mechanism in climbing beetles}}.
\newblock {\emph{\JournalTitle{J. R. Soc. Interface}}}
  \textbf{\bibinfo{volume}{14}}, \bibinfo{pages}{20170134}
  (\bibinfo{year}{2017}).

\bibitem{Lee2019}
\bibinfo{author}{Lee, S.~H.}, \bibinfo{author}{Song, H.~W.},
  \bibinfo{author}{Kang, B.~S.} \& \bibinfo{author}{Kwak, M.~K.}
\newblock \bibinfo{journal}{\bibinfo{title}{Remora-inspired reversible adhesive
  for underwater applications}}.
\newblock {\emph{\JournalTitle{ACS Appl. Mater. Interfaces}}}
  \textbf{\bibinfo{volume}{11}}, \bibinfo{pages}{47571--47576}
  (\bibinfo{year}{2019}).

\bibitem{Chuang2017}
\bibinfo{author}{Chuang, Y.-C.}, \bibinfo{author}{Chang, H.-K.},
  \bibinfo{author}{Liu, G.-L.} \& \bibinfo{author}{Chen, P.-Y.}
\newblock \bibinfo{journal}{\bibinfo{title}{Climbing upstream: Multi-scale
  structural characterization and underwater adhesion of the pulin river loach
  (sinogastromyzon puliensis)}}.
\newblock {\emph{\JournalTitle{J. Mech. Behav. Biomed. Mater.}}}
  \textbf{\bibinfo{volume}{73}}, \bibinfo{pages}{76--85}
  (\bibinfo{year}{2017}).

\bibitem{Ditsche2010}
\bibinfo{author}{Ditsche-Kuru, P.}, \bibinfo{author}{Koop, J.} \&
  \bibinfo{author}{Gorb, S.}
\newblock \bibinfo{journal}{\bibinfo{title}{Underwater attachment in current:
  the role of setose attachment structures on the gills of the mayfly larvae
  epeorus assimilis (ephemeroptera, heptageniidae)}}.
\newblock {\emph{\JournalTitle{J. Exp. Biol.}}} \textbf{\bibinfo{volume}{213}},
  \bibinfo{pages}{1950--1959} (\bibinfo{year}{2010}).

\bibitem{Kavanagh2001}
\bibinfo{author}{Kavanagh, C.~J.} \emph{et~al.}
\newblock \bibinfo{journal}{\bibinfo{title}{Variation in adhesion strength of
  balanus eburneus, crassostrea virginica and hydroides dianthus to
  fouling-release coatings}}.
\newblock {\emph{\JournalTitle{Biofouling}}} \textbf{\bibinfo{volume}{17}},
  \bibinfo{pages}{155--167} (\bibinfo{year}{2001}).

\bibitem{Rittschof2008}
\bibinfo{author}{Rittschof, D.} \emph{et~al.}
\newblock \bibinfo{journal}{\bibinfo{title}{Barnacle reattachment: a tool for
  studying barnacle adhesion}}.
\newblock {\emph{\JournalTitle{Biofouling}}} \textbf{\bibinfo{volume}{24}},
  \bibinfo{pages}{1--9} (\bibinfo{year}{2008}).

\bibitem{Holm2006}
\bibinfo{author}{Holm, E.~R.} \emph{et~al.}
\newblock \bibinfo{journal}{\bibinfo{title}{Interspecific variation in patterns
  of adhesion of marine fouling to silicone surfaces}}.
\newblock {\emph{\JournalTitle{Biofouling}}} \textbf{\bibinfo{volume}{22}},
  \bibinfo{pages}{233--243} (\bibinfo{year}{2006}).

\bibitem{Lin2007}
\bibinfo{author}{Lin, Q.} \emph{et~al.}
\newblock \bibinfo{journal}{\bibinfo{title}{Adhesion mechanisms of the mussel
  foot proteins mfp-1 and mfp-3}}.
\newblock {\emph{\JournalTitle{Proc. Natl. Acad. Sci. U. S. A}}}
  \textbf{\bibinfo{volume}{104}}, \bibinfo{pages}{3782--3786}
  (\bibinfo{year}{2007}).

\bibitem{George2018}
\bibinfo{author}{George, M.~N.}, \bibinfo{author}{Pedigo, B.} \&
  \bibinfo{author}{Carrington, E.}
\newblock \bibinfo{journal}{\bibinfo{title}{Hypoxia weakens mussel attachment
  by interrupting dopa cross-linking during adhesive plaque curing}}.
\newblock {\emph{\JournalTitle{J. R. Soc. Interface}}}
  \textbf{\bibinfo{volume}{15}}, \bibinfo{pages}{20180489}
  (\bibinfo{year}{2018}).

\bibitem{Priemel2021}
\bibinfo{author}{Priemel, T.} \emph{et~al.}
\newblock \bibinfo{journal}{\bibinfo{title}{Microfluidic-like fabrication of
  metal ion--cured bioadhesives by mussels}}.
\newblock {\emph{\JournalTitle{Science}}} \textbf{\bibinfo{volume}{374}},
  \bibinfo{pages}{206--211} (\bibinfo{year}{2021}).

\bibitem{Yu2013}
\bibinfo{author}{Yu, J.} \emph{et~al.}
\newblock \bibinfo{journal}{\bibinfo{title}{Adaptive hydrophobic and
  hydrophilic interactions of mussel foot proteins with organic thin films}}.
\newblock {\emph{\JournalTitle{Proc. Natl. Acad. Sci. U. S. A}}}
  \textbf{\bibinfo{volume}{110}}, \bibinfo{pages}{15680--15685}
  (\bibinfo{year}{2013}).

\bibitem{Lee2006}
\bibinfo{author}{Lee, H.}, \bibinfo{author}{Scherer, N.~F.} \&
  \bibinfo{author}{Messersmith, P.~B.}
\newblock \bibinfo{journal}{\bibinfo{title}{Single-molecule mechanics of mussel
  adhesion}}.
\newblock {\emph{\JournalTitle{Proc. Natl. Acad. Sci. U. S. A}}}
  \textbf{\bibinfo{volume}{103}}, \bibinfo{pages}{12999--13003}
  (\bibinfo{year}{2006}).

\bibitem{Waite2017}
\bibinfo{author}{Waite, J.~H.}
\newblock \bibinfo{journal}{\bibinfo{title}{Mussel adhesion--essential
  footwork}}.
\newblock {\emph{\JournalTitle{J. Exp. Biol.}}} \textbf{\bibinfo{volume}{220}},
  \bibinfo{pages}{517--530} (\bibinfo{year}{2017}).

\bibitem{Filippidi2015}
\bibinfo{author}{Filippidi, E.} \emph{et~al.}
\newblock \bibinfo{journal}{\bibinfo{title}{The microscopic network structure
  of mussel (mytilus) adhesive plaques}}.
\newblock {\emph{\JournalTitle{J. R. Soc. Interface}}}
  \textbf{\bibinfo{volume}{12}}, \bibinfo{pages}{20150827}
  (\bibinfo{year}{2015}).

\bibitem{Qin2013}
\bibinfo{author}{Qin, Z.} \& \bibinfo{author}{Buehler, M.~J.}
\newblock \bibinfo{journal}{\bibinfo{title}{Impact tolerance in mussel thread
  networks by heterogeneous material distribution}}.
\newblock {\emph{\JournalTitle{Nat. Commun.}}} \textbf{\bibinfo{volume}{4}},
  \bibinfo{pages}{1--8} (\bibinfo{year}{2013}).

\bibitem{Valois2019}
\bibinfo{author}{Valois, E.}, \bibinfo{author}{Hoffman, C.},
  \bibinfo{author}{Demartini, D.~G.} \& \bibinfo{author}{Waite, J.~H.}
\newblock \bibinfo{journal}{\bibinfo{title}{The thiol-rich interlayer in the
  shell/core architecture of mussel byssal threads}}.
\newblock {\emph{\JournalTitle{Langmuir}}} \textbf{\bibinfo{volume}{35}},
  \bibinfo{pages}{15985--15991} (\bibinfo{year}{2019}).

\bibitem{Jehle2020}
\bibinfo{author}{Jehle, F.} \emph{et~al.}
\newblock \bibinfo{journal}{\bibinfo{title}{Hierarchically-structured
  metalloprotein composite coatings biofabricated from co-existing condensed
  liquid phases}}.
\newblock {\emph{\JournalTitle{Nat. Commun.}}} \textbf{\bibinfo{volume}{11}},
  \bibinfo{pages}{1--9} (\bibinfo{year}{2020}).

\bibitem{Harrington2010}
\bibinfo{author}{Harrington, M.~J.}, \bibinfo{author}{Masic, A.},
  \bibinfo{author}{Holten-Andersen, N.}, \bibinfo{author}{Waite, J.~H.} \&
  \bibinfo{author}{Fratzl, P.}
\newblock \bibinfo{journal}{\bibinfo{title}{Iron-clad fibers: a metal-based
  biological strategy for hard flexible coatings}}.
\newblock {\emph{\JournalTitle{Science}}} \textbf{\bibinfo{volume}{328}},
  \bibinfo{pages}{216--220} (\bibinfo{year}{2010}).

\bibitem{Bernstein2020}
\bibinfo{author}{Bernstein, J.~H.}, \bibinfo{author}{Filippidi, E.},
  \bibinfo{author}{Waite, J.~H.} \& \bibinfo{author}{Valentine, M.~T.}
\newblock \bibinfo{journal}{\bibinfo{title}{Effects of sea water ph on marine
  mussel plaque maturation}}.
\newblock {\emph{\JournalTitle{Soft Matter}}} \textbf{\bibinfo{volume}{16}},
  \bibinfo{pages}{9339--9346} (\bibinfo{year}{2020}).

\bibitem{Chisca2018}
\bibinfo{author}{Chisca, S.}, \bibinfo{author}{Musteata, V.-E.},
  \bibinfo{author}{Sougrat, R.}, \bibinfo{author}{Behzad, A.~R.} \&
  \bibinfo{author}{Nunes, S.~P.}
\newblock \bibinfo{journal}{\bibinfo{title}{Artificial 3d hierarchical and
  isotropic porous polymeric materials}}.
\newblock {\emph{\JournalTitle{Sci. Adv.}}} \textbf{\bibinfo{volume}{4}},
  \bibinfo{pages}{eaat0713} (\bibinfo{year}{2018}).

\bibitem{Yang2022}
\bibinfo{author}{Yang, T.} \emph{et~al.}
\newblock \bibinfo{journal}{\bibinfo{title}{High strength and damage-tolerance
  in echinoderm stereom as a natural bicontinuous ceramic cellular solid}}.
\newblock {\emph{\JournalTitle{Nat. Commun.}}} \textbf{\bibinfo{volume}{13}},
  \bibinfo{pages}{1--12} (\bibinfo{year}{2022}).

\bibitem{Shi2021}
\bibinfo{author}{Shi, S.}, \bibinfo{author}{Li, Y.}, \bibinfo{author}{Ngo-Dinh,
  B.-N.}, \bibinfo{author}{Markmann, J.} \& \bibinfo{author}{Weissm{\"u}ller,
  J.}
\newblock \bibinfo{journal}{\bibinfo{title}{Scaling behavior of stiffness and
  strength of hierarchical network nanomaterials}}.
\newblock {\emph{\JournalTitle{Science}}} \textbf{\bibinfo{volume}{371}},
  \bibinfo{pages}{1026--1033} (\bibinfo{year}{2021}).

\bibitem{Pham2019}
\bibinfo{author}{Pham, M.-S.}, \bibinfo{author}{Liu, C.},
  \bibinfo{author}{Todd, I.} \& \bibinfo{author}{Lertthanasarn, J.}
\newblock \bibinfo{journal}{\bibinfo{title}{Damage-tolerant architected
  materials inspired by crystal microstructure}}.
\newblock {\emph{\JournalTitle{Nature}}} \textbf{\bibinfo{volume}{565}},
  \bibinfo{pages}{305--311} (\bibinfo{year}{2019}).

\bibitem{Liu2021}
\bibinfo{author}{Liu, C.}, \bibinfo{author}{Lertthanasarn, J.} \&
  \bibinfo{author}{Pham, M.-S.}
\newblock \bibinfo{journal}{\bibinfo{title}{The origin of the boundary
  strengthening in polycrystal-inspired architected materials}}.
\newblock {\emph{\JournalTitle{Nat. Commun.}}} \textbf{\bibinfo{volume}{12}},
  \bibinfo{pages}{1--10} (\bibinfo{year}{2021}).

\bibitem{Shaikeea2022}
\bibinfo{author}{Shaikeea, A. J.~D.}, \bibinfo{author}{Cui, H.},
  \bibinfo{author}{O’Masta, M.}, \bibinfo{author}{Zheng, X.~R.} \&
  \bibinfo{author}{Deshpande, V.~S.}
\newblock \bibinfo{journal}{\bibinfo{title}{The toughness of mechanical
  metamaterials}}.
\newblock {\emph{\JournalTitle{Nat. Mater.}}} \textbf{\bibinfo{volume}{21}},
  \bibinfo{pages}{297--304} (\bibinfo{year}{2022}).

\bibitem{Bhuwal2023}
\bibinfo{author}{Bhuwal, A.~S.}, \bibinfo{author}{Pang, Y.},
  \bibinfo{author}{Ashcroft, I.}, \bibinfo{author}{Sun, W.} \&
  \bibinfo{author}{Liu, T.}
\newblock \bibinfo{journal}{\bibinfo{title}{Discovery of quasi-disordered truss
  metamaterials inspired by natural cellular materials}}.
\newblock {\emph{\JournalTitle{J. Mech. Phys. Solids}}} \bibinfo{pages}{105294}
  (\bibinfo{year}{2023}).

\bibitem{Yang20222}
\bibinfo{author}{Yang, T.} \emph{et~al.}
\newblock \bibinfo{journal}{\bibinfo{title}{A damage-tolerant, dual-scale,
  single-crystalline microlattice in the knobby starfish, protoreaster
  nodosus}}.
\newblock {\emph{\JournalTitle{Science}}} \textbf{\bibinfo{volume}{375}},
  \bibinfo{pages}{647--652} (\bibinfo{year}{2022}).

\bibitem{Aldred2007}
\bibinfo{author}{Aldred, N.}, \bibinfo{author}{Wills, T.},
  \bibinfo{author}{Williams, D.} \& \bibinfo{author}{Clare, A.}
\newblock \bibinfo{journal}{\bibinfo{title}{Tensile and dynamic mechanical
  analysis of the distal portion of mussel (mytilus edulis) byssal threads}}.
\newblock {\emph{\JournalTitle{J. R. Soc. Interface}}}
  \textbf{\bibinfo{volume}{4}}, \bibinfo{pages}{1159--1167}
  (\bibinfo{year}{2007}).

\bibitem{Xu2019}
\bibinfo{author}{Xu, Q.} \emph{et~al.}
\newblock \bibinfo{journal}{\bibinfo{title}{Metal coordination-mediated
  functional grading and self-healing in mussel byssus cuticle}}.
\newblock {\emph{\JournalTitle{Adv. Sci.}}} \textbf{\bibinfo{volume}{6}},
  \bibinfo{pages}{1902043} (\bibinfo{year}{2019}).

\bibitem{Rao2022}
\bibinfo{author}{Renner-Rao, M.} \emph{et~al.}
\newblock \bibinfo{journal}{\bibinfo{title}{Mussels fabricate porous glues via
  multiphase liquid--liquid phase separation of multiprotein condensates}}.
\newblock {\emph{\JournalTitle{ACS nano}}} \textbf{\bibinfo{volume}{16}},
  \bibinfo{pages}{20877--20890} (\bibinfo{year}{2022}).

\bibitem{Desmond2015}
\bibinfo{author}{Desmond, K.~W.}, \bibinfo{author}{Zacchia, N.~A.},
  \bibinfo{author}{Waite, J.~H.} \& \bibinfo{author}{Valentine, M.~T.}
\newblock \bibinfo{journal}{\bibinfo{title}{Dynamics of mussel plaque
  detachment}}.
\newblock {\emph{\JournalTitle{Soft matter}}} \textbf{\bibinfo{volume}{11}},
  \bibinfo{pages}{6832--6839} (\bibinfo{year}{2015}).

\bibitem{Kwon2021}
\bibinfo{author}{Kwon, Y.}, \bibinfo{author}{Bernstein, J.~H.},
  \bibinfo{author}{Cohen, N.} \& \bibinfo{author}{Valentine, M.~T.}
\newblock \bibinfo{journal}{\bibinfo{title}{On-demand manufacturing
  capabilities of mussels enable robust adhesion to geometrically complex
  surfaces}}.
\newblock {\emph{\JournalTitle{ACS Biomater. Sci. Eng.}}}
  \textbf{\bibinfo{volume}{7}}, \bibinfo{pages}{5099--5106}
  (\bibinfo{year}{2021}).

\bibitem{Amini2017}
\bibinfo{author}{Amini, S.} \emph{et~al.}
\newblock \bibinfo{journal}{\bibinfo{title}{Preventing mussel adhesion using
  lubricant-infused materials}}.
\newblock {\emph{\JournalTitle{Science}}} \textbf{\bibinfo{volume}{357}},
  \bibinfo{pages}{668--673} (\bibinfo{year}{2017}).

\bibitem{Kang2016}
\bibinfo{author}{Kang, T.} \emph{et~al.}
\newblock \bibinfo{journal}{\bibinfo{title}{Mussel-inspired anchoring of
  polymer loops that provide superior surface lubrication and antifouling
  properties}}.
\newblock {\emph{\JournalTitle{Acs Nano}}} \textbf{\bibinfo{volume}{10}},
  \bibinfo{pages}{930--937} (\bibinfo{year}{2016}).

\bibitem{Carrington2004}
\bibinfo{author}{Carrington, E.} \& \bibinfo{author}{Gosline, J.~M.}
\newblock \bibinfo{journal}{\bibinfo{title}{Mechanical design of mussel byssus:
  load cycle and strain rate dependence}}.
\newblock {\emph{\JournalTitle{Amal. Malacol. Bull.}}}
  \textbf{\bibinfo{volume}{18}}, \bibinfo{pages}{135--142}
  (\bibinfo{year}{2004}).

\bibitem{Wilhelm2017}
\bibinfo{author}{Wilhelm, M.~H.}, \bibinfo{author}{Filippidi, E.},
  \bibinfo{author}{Waite, J.~H.} \& \bibinfo{author}{Valentine, M.~T.}
\newblock \bibinfo{journal}{\bibinfo{title}{Influence of multi-cycle loading on
  the structure and mechanics of marine mussel plaques}}.
\newblock {\emph{\JournalTitle{Soft Matter}}} \textbf{\bibinfo{volume}{13}},
  \bibinfo{pages}{7381--7388} (\bibinfo{year}{2017}).

\bibitem{Bertoldi2007}
\bibinfo{author}{Bertoldi, K.} \& \bibinfo{author}{Boyce, M.~C.}
\newblock \bibinfo{journal}{\bibinfo{title}{Mechanics of the hysteretic large
  strain behavior of mussel byssus threads}}.
\newblock {\emph{\JournalTitle{J. Mater. Sci.}}} \textbf{\bibinfo{volume}{42}},
  \bibinfo{pages}{8943--8956} (\bibinfo{year}{2007}).

\bibitem{Choi2021}
\bibinfo{author}{Choi, J.}, \bibinfo{author}{Lee, S.}, \bibinfo{author}{Ohkawa,
  K.} \& \bibinfo{author}{Hwang, D.~S.}
\newblock \bibinfo{journal}{\bibinfo{title}{Counterplotting the
  mechanosensing-based fouling mechanism of mussels against fouling}}.
\newblock {\emph{\JournalTitle{ACS nano}}} \textbf{\bibinfo{volume}{15}},
  \bibinfo{pages}{18566--18579} (\bibinfo{year}{2021}).

\bibitem{Liu2009}
\bibinfo{author}{Liu, M.}, \bibinfo{author}{Sun, J.}, \bibinfo{author}{Sun,
  Y.}, \bibinfo{author}{Bock, C.} \& \bibinfo{author}{Chen, Q.}
\newblock \bibinfo{journal}{\bibinfo{title}{Thickness-dependent mechanical
  properties of polydimethylsiloxane membranes}}.
\newblock {\emph{\JournalTitle{J. Micromech. Microeng}}}
  \textbf{\bibinfo{volume}{19}}, \bibinfo{pages}{035028}
  (\bibinfo{year}{2009}).

\bibitem{Ogden1972}
\bibinfo{author}{Ogden, R.~W.}
\newblock \bibinfo{journal}{\bibinfo{title}{Large deformation isotropic
  elasticity--on the correlation of theory and experiment for incompressible
  rubberlike solids}}.
\newblock {\emph{\JournalTitle{Proc R Soc Lond A Math Phys Sci.}}}
  \textbf{\bibinfo{volume}{326}}, \bibinfo{pages}{565--584}
  (\bibinfo{year}{1972}).

\bibitem{Luke2020}
\bibinfo{author}{Gockowski, L.~F.} \emph{et~al.}
\newblock \bibinfo{journal}{\bibinfo{title}{Engineering crack tortuosity in
  printed polymer--polymer composites through ordered pores}}.
\newblock {\emph{\JournalTitle{Materials Horizons}}}
  \textbf{\bibinfo{volume}{7}}, \bibinfo{pages}{1854--1860}
  (\bibinfo{year}{2020}).

\bibitem{Borys2016}
\bibinfo{author}{Drach, B.}, \bibinfo{author}{Tsukrov, I.} \&
  \bibinfo{author}{Trofimov, A.}
\newblock \bibinfo{journal}{\bibinfo{title}{Comparison of full field and single
  pore approaches to homogenization of linearly elastic materials with pores of
  regular and irregular shapes}}.
\newblock {\emph{\JournalTitle{Int J Solids Struct}}}
  \textbf{\bibinfo{volume}{96}}, \bibinfo{pages}{48--63}
  (\bibinfo{year}{2016}).

\bibitem{Cao2019}
\bibinfo{author}{Cao, S.}, \bibinfo{author}{Liu, T.}, \bibinfo{author}{Jones,
  A.} \& \bibinfo{author}{Tizani, W.}
\newblock \bibinfo{journal}{\bibinfo{title}{Particle reinforced thermoplastic
  foams under quasi-static compression}}.
\newblock {\emph{\JournalTitle{Mechanics of Materials}}}
  \textbf{\bibinfo{volume}{136}}, \bibinfo{pages}{103081}
  (\bibinfo{year}{2019}).

\bibitem{Valois2020}
\bibinfo{author}{Valois, E.}, \bibinfo{author}{Mirshafian, R.} \&
  \bibinfo{author}{Waite, J.~H.}
\newblock \bibinfo{journal}{\bibinfo{title}{Phase-dependent redox insulation in
  mussel adhesion}}.
\newblock {\emph{\JournalTitle{Sci. Adv.}}} \textbf{\bibinfo{volume}{6}},
  \bibinfo{pages}{eaaz6486} (\bibinfo{year}{2020}).

\bibitem{Camanho2002}
\bibinfo{author}{Camanho, P.~P.} \& \bibinfo{author}{D{\'a}vila, C.~G.}
\newblock \bibinfo{journal}{\bibinfo{title}{Mixed-mode decohesion finite
  elements for the simulation of delamination in composite materials}}.
\newblock {\emph{\JournalTitle{NASA/TM-2002–211737}}} \bibinfo{pages}{1--37}
  (\bibinfo{year}{2002}).

\bibitem{Pritchard2013}
\bibinfo{author}{Pritchard, R.~H.}, \bibinfo{author}{Lava, P.},
  \bibinfo{author}{Debruyne, D.} \& \bibinfo{author}{Terentjev, E.~M.}
\newblock \bibinfo{journal}{\bibinfo{title}{Precise determination of the
  poisson ratio in soft materials with 2d digital image correlation}}.
\newblock {\emph{\JournalTitle{Soft Matter}}} \textbf{\bibinfo{volume}{9}},
  \bibinfo{pages}{6037--6045} (\bibinfo{year}{2013}).

\bibitem{Smeathers1979}
\bibinfo{author}{Smeathers, J.} \& \bibinfo{author}{Vincent, J.}
\newblock \bibinfo{journal}{\bibinfo{title}{Mechanical properties of mussel
  byssus threads}}.
\newblock {\emph{\JournalTitle{J. Molluscan Stud.}}}
  \textbf{\bibinfo{volume}{45}}, \bibinfo{pages}{219--230}
  (\bibinfo{year}{1979}).

\bibitem{Priemel2017}
\bibinfo{author}{Priemel, T.}, \bibinfo{author}{Degtyar, E.},
  \bibinfo{author}{Dean, M.~N.} \& \bibinfo{author}{Harrington, M.~J.}
\newblock \bibinfo{journal}{\bibinfo{title}{Rapid self-assembly of complex
  biomolecular architectures during mussel byssus biofabrication}}.
\newblock {\emph{\JournalTitle{Nat. Commun.}}} \textbf{\bibinfo{volume}{8}},
  \bibinfo{pages}{1--12} (\bibinfo{year}{2017}).

\bibitem{Nordestgaard1985}
\bibinfo{author}{Nordestgaard, B.} \& \bibinfo{author}{Rostgaard, J.}
\newblock \bibinfo{journal}{\bibinfo{title}{Critical-point drying versus freeze
  drying for scanning electron microscopy: a quantitative and qualitative study
  on isolated hepatocytes}}.
\newblock {\emph{\JournalTitle{J. Microsc.}}} \textbf{\bibinfo{volume}{137}},
  \bibinfo{pages}{189--207} (\bibinfo{year}{1985}).

\bibitem{Bray1993}
\bibinfo{author}{Bray, D.}, \bibinfo{author}{Bagu, J.} \&
  \bibinfo{author}{Koegler, P.}
\newblock \bibinfo{journal}{\bibinfo{title}{Comparison of hexamethyldisilazane
  (hmds), peldri ii, and critical-point drying methods for scanning electron
  microscopy of biological specimens}}.
\newblock {\emph{\JournalTitle{Microsc. Res. Tech.}}}
  \textbf{\bibinfo{volume}{26}}, \bibinfo{pages}{489--495}
  (\bibinfo{year}{1993}).

\bibitem{ASTMD412}
\bibinfo{author}{{ASTM D412-16}}.
\newblock \bibinfo{journal}{\bibinfo{title}{Standard test methods for
  vulcanized rubber and thermoplastic elastomers-tension}}.
\newblock {\emph{\JournalTitle{ASTM International, West Conshohocken, PA.}}}
  (\bibinfo{year}{2016}).

\bibitem{Moon2013}
\bibinfo{author}{Moon~Jeong, S.}, \bibinfo{author}{Song, S.},
  \bibinfo{author}{Lee, S.-K.} \& \bibinfo{author}{Choi, B.}
\newblock \bibinfo{journal}{\bibinfo{title}{Mechanically driven light-generator
  with high durability}}.
\newblock {\emph{\JournalTitle{Appl. Phys. Lett.}}}
  \textbf{\bibinfo{volume}{102}}, \bibinfo{pages}{051110}
  (\bibinfo{year}{2013}).

\bibitem{Park2019}
\bibinfo{author}{Park, H.-J.} \emph{et~al.}
\newblock \bibinfo{journal}{\bibinfo{title}{Self-powered motion-driven
  triboelectric electroluminescence textile system}}.
\newblock {\emph{\JournalTitle{ACS Appl. Mater. Interfaces}}}
  \textbf{\bibinfo{volume}{11}}, \bibinfo{pages}{5200--5207}
  (\bibinfo{year}{2019}).

\bibitem{Stark2013}
\bibinfo{author}{Stark, A.~Y.} \emph{et~al.}
\newblock \bibinfo{journal}{\bibinfo{title}{Surface wettability plays a
  significant role in gecko adhesion underwater}}.
\newblock {\emph{\JournalTitle{Proc. Natl. Acad. Sci. U.S.A}}}
  \textbf{\bibinfo{volume}{110}}, \bibinfo{pages}{6340--6345}
  (\bibinfo{year}{2013}).

\bibitem{DICe2015}
\bibinfo{author}{Turner, D.}
\newblock \bibinfo{journal}{\bibinfo{title}{Digital image correlation engine
  (dice) reference manual}}.
\newblock {\emph{\JournalTitle{Sandia Report, SAND2015-10606 O}}}
  (\bibinfo{year}{2015}).

\bibitem{Phillip2015}
\bibinfo{author}{Reu, P.}
\newblock \bibinfo{journal}{\bibinfo{title}{All about speckles: Speckle
  density}}.
\newblock {\emph{\JournalTitle{Exp. Tech.}}} \textbf{\bibinfo{volume}{39}},
  \bibinfo{pages}{1--2} (\bibinfo{year}{2015}).

\bibitem{Pang2020}
\bibinfo{author}{Pang, Y.}, \bibinfo{author}{Chen, B.~K.}, \bibinfo{author}{Yu,
  S.~F.} \& \bibinfo{author}{Lingamanaik, S.~N.}
\newblock \bibinfo{journal}{\bibinfo{title}{Enhanced laser speckle optical
  sensor for in situ strain sensing and structural health monitoring}}.
\newblock {\emph{\JournalTitle{Opt. Lett.}}} \textbf{\bibinfo{volume}{45}},
  \bibinfo{pages}{2331--2334} (\bibinfo{year}{2020}).

\bibitem{Su2019}
\bibinfo{author}{Su, Y.} \emph{et~al.}
\newblock \bibinfo{journal}{\bibinfo{title}{Theoretical analysis on performance
  of digital speckle pattern: uniqueness, accuracy, precision, and spatial
  resolution}}.
\newblock {\emph{\JournalTitle{Opt. Express}}} \textbf{\bibinfo{volume}{27}},
  \bibinfo{pages}{22439--22474} (\bibinfo{year}{2019}).

\bibitem{Harrington2018}
\bibinfo{author}{Harrington, M.~J.}, \bibinfo{author}{Jehle, F.} \&
  \bibinfo{author}{Priemel, T.}
\newblock \bibinfo{journal}{\bibinfo{title}{Mussel byssus structure-function
  and fabrication as inspiration for biotechnological production of advanced
  materials}}.
\newblock {\emph{\JournalTitle{Biotechnol. J.}}} \textbf{\bibinfo{volume}{13}},
  \bibinfo{pages}{1800133} (\bibinfo{year}{2018}).

\end{thebibliography}

\section*{Acknowledgements}
This work was funded by the Leverhulme Trust Research Grant Scheme, UK (No. RPG-2020-235) and NanoCAT Research Grant, the University of Nottingham.

\section*{Author contributions}
Y. P. carried out the laboratory tests, numerical simulation, microstructure characterization and wrote the first draft. T. L. conceived, designed and supervised this study and corrected the manuscript. All authors discussed the results and commented on the manuscript.

\section*{Competing interests}
The authors declare no competing interests.

\end{document}